\documentclass[secnumarabic,twocolumn,aps,pre,showpacs,showkey,superscriptaddress,prl,floatfix]{revtex4-1}

\usepackage{amsmath,amssymb,amsthm}
\usepackage[capbesideposition=right]{floatrow}

\usepackage{hyperref}
\usepackage{amsthm}
\usepackage{graphicx}
\usepackage{subfigure}
\usepackage{color,bbm,caption}
\usepackage[normalem]{ulem}

\graphicspath{{../Figures/}}

\newcommand{\R}{\mathbb{R}}

\newcommand{\erf}{\mathrm{erf}}
\def \E {\mathbb{E}}

\newcommand{\eps}{\varepsilon}

\theoremstyle{plain}

\theoremstyle{definition}

\renewcommand{\sout}[1]{}

\begin{document}

\title{Noise-induced synchronization and anti-resonance in excitable systems; Implications for information processing in Parkinson's Disease and Deep Brain Stimulation}
\author{Jonathan D. Touboul}
\email[]{jtouboul@brandeis.edu}
\affiliation{Department of Mathematics and Volen Center for Complex Systems, Brandeis University.}
\author{Charlotte Piette}
\affiliation{Department of Mathematics and Volen Center for Complex Systems, Brandeis University.}
\affiliation{CIRB - Coll\`ege de France.}
\author{Laurent Venance}
\affiliation{CIRB - Coll\`ege de France.}
\author{G. Bard Ermentrout}
\affiliation{Department of Mathematics, University of Pittsburgh.}

\date{\today}%

\begin{abstract}
We study the statistical physics of a surprising phenomenon arising in large networks of excitable elements in response to noise: while at low noise, solutions remain in the vicinity of the resting state and large-noise solutions show asynchronous activity, the network displays orderly, perfectly synchronized periodic responses at intermediate levels of noise. This noise-induced synchronization, distinct from classical stochastic resonances, is fundamentally collective in nature. Indeed, we show that, for noise and coupling within specific ranges, an asymmetry in the transition rates between a resting and an excited regime progressively builds up, leading to an increase in the fraction of excited neurons eventually triggering a chain reaction associated with a macroscopic synchronized excursion and a collective return to rest where this process starts afresh, thus yielding the observed periodic synchronized oscillations. We further uncover a novel anti-resonance phenomenon in this regime: noise-induced synchronized oscillations disappear when the system is driven by periodic stimulation with frequency within a specific range (high relative to the spontaneous activity). In that anti-resonance regime, the system is optimal for measures of information capacity. This observation provides a new hypothesis accounting for the efficiency of high-frequency stimulation therapies, known as Deep Brain Stimulation, in Parkinson's disease, a neurodegenerative disease characterized by an increased synchronization of brain motor circuits. We further discuss the universality of these phenomena in the class of stochastic networks of excitable elements with confining coupling, and illustrate this universality by analyzing various classical models of neuronal networks. Altogether, these results uncover some universal mechanisms supporting a regularizing impact of noise in excitable systems, reveal a novel anti-resonance phenomenon in these systems, and propose a new hypothesis for the efficiency of high-frequency stimulation in Parkinson's disease. 
\end{abstract} 

\maketitle


Coupled systems of excitable elements subject to noise are commonly used to model natural and physical phenomena. They describe in particular laser emission~\cite{goulding}, chemical reactions where noise reportedly supports traveling waves~\cite{Kadar}, climate dynamics~\cite{levin}, cardiac tissue and other physiological processes~\cite{glass}, gene networks where excitability in the presence of noise was suggested as a possible mechanism for transient cellular differentiation~\cite{suel}, and neurons and ion channels~\cite{ermentrout-terman:10b} (see~\cite{lindner:04} for a review).

Excitable systems have in common the existence of a \emph{rest} state, a globally attractive fixed point when the system is unperturbed, and two typical responses to perturbations: small perturbations result in small amplitude responses, while sufficiently strong perturbations (bringing the system to cross a \emph{separatrix}) lead to a long excursion (a spike) through an \emph{excited} state, followed by a return to rest after a \emph{refractory period} during which the system essentially cannot be excited. Single excitable elements subject to noise have been widely studied theoretically and experimentally, and various stochastic resonances such as coherence resonance or self-induced stochastic resonance (SISR), were identified~\cite{lindner:04}. The latter phenomenon is associated with maximally periodic responses of a single excitable unit in response to small noise, as thoroughly described for the FitzHugh-Nagumo model in~\cite{deville2005two}, in a limit of timescale separation and vanishing noise with a common scaling. While subtle and analyzed in asymptotic regimes, these phenomena are not purely abstract, and their footprint was evoked in various natural phenomena particularly in neural systems~\cite{levin1996broadband,collins1996noise,wiesenfeld1995stochastic}. How these phenomena scale up in large networks of excitable elements and with larger noise levels compatible with typical fluctuations in natural systems remains a largely open problem. 

Synchronized oscillations constitute a significant macroscopic state in networks of excitable systems. In brain for instance, rhythmic macroscopic activity (a hallmark of collective neuronal synchronization) was observed in a variety of species, in various brain areas, and across a wide range of frequencies~\cite{buszaki:06}, and, in mammals, are reportedly related to various cognitive processes such as memory, attention and sleep~\cite{wang:10}. Impairments in synchronous activity are also observed in several pathologies: abnormally high synchrony is reported in epilepsy or Parkinson's disease, low synchrony in Alzheimer's disease, and altered oscillatory patterns in schizophrenia~\cite{uhlhaas-singer:06}. These regular behaviors emerge despite the presence, in all physical and biological systems, of multifarious noisy fluctuations (see e.g.~\cite{aldo-faisal:08} for a review of sources of noise in the brain), that often have a significant impact on the dynamics. Important progress has been done to understand the synchronization of oscillators (dynamical systems that oscillate intrinsically), in particular in the frame of the Kuramoto model. An abundant literature has characterized how these oscillators can generate coherent behaviors in the presence of noise, heterogeneity in the intrinsic frequencies, graph structure of interactions~\cite{acebron,strogatz,bag}, and even, particularly relevant in the present context, how noise and periodic input can shape synchronization of non-identical oscillators~\cite{lai-porter}. Excitable systems, such as neurons, are not intrinsic oscillators: in the absence of input, they stabilize at a fixed point, and thus synchronization of excitable elements likely relies on mechanisms in large part distinct from those of coupled oscillators. 

This paper investigates a surprising and somewhat paradoxical regularizing impact of noise in large-scale networks of excitable elements: as noise is progressively increased, the network shows a sudden transition from stationary, low-amplitude fluctuations (\emph{clamped regime}) to a massive synchronization, yielding coherent, high amplitude and periodic macroscopic oscillations (\emph{noise-induced oscillations regime}), that progressively desynchronize as noise is further increased  until an \emph{aynchronous regime} is reached where randomness overwhelms collective effects. This phenomenon, predicted in abstract nonlinear diffusions of mean-field type more than thirty years ago~\cite{scheutzow1985noise,scheutzow1985some}, was studied in a specific neuronal network~\cite{touboul-hermann-faugeras:11} using Gaussian properties of the solutions of that particular system, and has been the topic of renewed interest recently in mathematics~\cite{lucon,quininao-touboul:18}. Those mathematical approaches, relying on fine investigations of the solutions of a mean-field equation, unfortunately do not describe the mechanisms supporting the emergence of these oscillations. In the limit of small noise and high excitability (resting state lying in the vicinity of the excitability separatrix), Zaks and collaborators proposed to use a moment expansion and closure based on the assumption that solutions are approximately Gaussian~\cite{zaks,lindner:04}, and showed how increasing noise could lead to periodic behaviors. This Gaussian approximation however does not extend beyond small noise; as noise is increased, the Gaussian approximation breaks down, and we will exhibit the emergence of a bimodal distribution crucial in the generation of noise-induced oscillations. We will indeed show that an increasing fraction of neurons tend to enter the excited state, forming a second peak in the distribution that will act as a key driver for the collective synchronization. 

Altogether, despite these previous works, little remains known about the microscopic mechanisms supporting noise-induced oscillations in excitable systems, and progress in numerical analysis or mathematical analyses of mean-field equations does not allow for addressing those mechanisms. We develop here a fine scale analysis of individual trajectories, in a stochastic electrically coupled FitzHugh-Nagumo network, to unravel the microscopic mechanisms at play. Electrical coupling was indeed shown to favor the emergence of synchronized activity in biological neural networks~\cite{connors:04,connors:17,coulon:17,alcami:19}~\footnote{{Electrical synapses are widely observed throughout the mammalian and invertebrate nervous system~\cite{bennett2004electrical}, but may be less frequent than chemical synapses in some neural networks. This choice of coupling simplifies our theoretical derivations. The phenomena reported are however more general than neural networks with electrical coupling, and when exploring universality, we will show that \emph{confining} coupling leads to synchronization, and such couplings include in particular neural networks with chemical coupling and excitatory and inhibitory populations.}}; it is also a particularly simple mathematical model allowing an in-depth study. This analysis will  highlight the respective roles of coupling levels, noise intensity and excitability in the emergence of oscillations due to noise.  Going further, these statistical physics mechanisms will lead us to uncover a novel phenomenon of coupled excitable systems in the noise-induced synchronization regime. We will show that a periodic forcing of the system at high frequency (relative to the spontaneous noise-induced oscillation frequency) can prevent synchronization, and set the system in a regime where it is able to optimally process information. Our analysis of the mechanisms leading to synchronization in turn will allow us to identify the critical requirements underlying these behaviors, highlight their universality for excitable networks with confining interactions, and to conjecture the type of transition to synchrony occurring  as noise or connectivity are varied.

These phenomena may have multiple applications. We particularly explore here their implications in the context of Parkinson's disease and its treatment, and will use vocabulary and concepts from neuroscience throughout the paper. Parkinson's disease is a neurodegenerative disorder classically associated with dramatic motor and cognitive symptoms, and with a pathological elevation of oscillations in the beta frequency band -- 13-30 Hz -- in the basal ganglia and motor cortex in particular~\cite{hammond:07,mcgregor:19}. Among the variety of factors contributing to these oscillations, studies invoked an elevated excitability of neurons~\cite{payroux:04,chen,degos} and increased electrical coupling~\cite{schwab2014pallidal,phookan2015gap}, two elements that we will see are important in the emergence of noise-induced oscillations in our models. Deep Brain Stimulation (DBS), an efficient symptomatic treatment of Parkinson's disease, consists of stimulating periodically at a high-frequency (130 Hz) the subthalamic nucleus in the basal ganglia~\cite{benabid1987combined,perlmutter2006deep,bain2009deep,limousin:95,ashkan,limousin:95}, and leads to a remarkable reduction of motor symptoms and abnormal beta-band synchronization. Yet the mechanisms of action of DBS have remained elusive, and stimulation parameters are largely tuned heuristically and sometimes need to be readjusted following a subsequent emergence of neuropsychological symptoms~\cite{kumar1998double,rodriguez2005bilateral,aum}. Computational models speculated two possible mechanisms of action of DBS~\cite{modolo}: by increasing inhibitory currents and altering inhibitory firing pattern~\cite{rubin2004high} or by depolarization blockade~\cite{breit:04,heida2009effectiveness,benazzouz2000mechanism}. Here, we demonstrate that anti-resonance in excitable networks could also serve as a hypothetical mechanism which, devoid of increased inhibition or excitation blockade, leaves the network highly responsive to stimuli and thus allows restoring cognitive processes. We will show that indeed, in the anti-resonance regime, the system displays optimal information processing capabilities, potentially joining clinical observations of DBS reportedly restoring motor and cognitive function in parkinsonian patients. 

The paper is organized as follows. Section~\ref{sec:NoiseInducedSync} introduces our reference neural network model, the electrically coupled FitzHugh-Nagumo network, and describes numerically the emergence of the noise-induced synchronization in this model. Section~\ref{sec:NoiseInducedModel} is devoted to deciphering the statistical physics mechanisms underpinning this synchronization, while section~\ref{sec:Desynchronization} unravels and analyzes the anti-resonance phenomenon and the associated information processing capabilities. We discuss the universality of these phenomena in section~\ref{sec:Universality}.

\section{Model and Noise-Induced Synchronization}\label{sec:NoiseInducedSync}
The electrically coupled network of FitzHugh-Nagumo neurons~\cite{fitzhugh:55,ermentrout-terman:10b} describes the dynamics of $n$ neurons through the equations:
\begin{equation}\label{eq:FhN}
\begin{cases}
dv^{i}_{t} = \Big(f(v^{i}_{t})-w^{i}_{t} + \frac{J}{n}\sum_{j=1}^{n} (v^{j}_{t}-v^{i}_{t}) + I(t)\Big)\,dt + \sigma dW^{i}_{t}\\
dw^{i}_{t} = \eps(b v^{i}_{t}-w^{i}_{t})\,dt
\end{cases}
\end{equation}
where $i\in \{1,\cdots, n\}$ denotes the neuron index, $v^{i}$ the associated voltage and $w^{i}$ the associated recovery (or adaptation) variable. The function $f$ is a cubic nonlinearity modeling the excitability of the cells, classically considered as:
\[f(v)=v(1-v)(v-a),\]
where $a>0$ controls the excitability, $J>0$ quantifies the coupling level, $I(t)$ is an input current, $\sigma$ the level of noise,  $(W^{i}_{t})$ are independent Brownian motions, the timescale ratio $\eps>0$ of adaptation compared to voltage is generally assumed small, and $b>0$ governs the coupling between voltage and adaptation. Each neuron isolated, satisfying the FitzHugh-Nagumo equations, is thus classically an excitable system. 

\begin{figure*}
	\includegraphics[width=\textwidth]{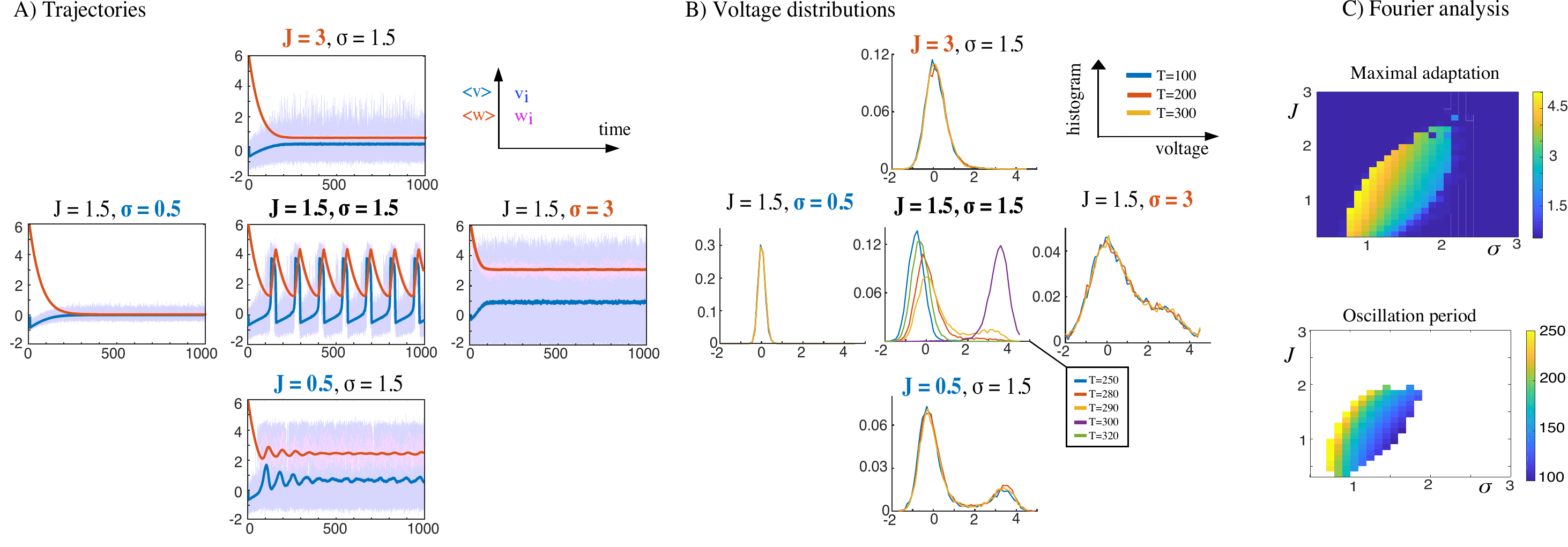}
	\caption{Noise induced synchronization in the FitzHugh-Nagumo model, with $n=4\,000$ neurons ($n=10\,000$ for histograms in panels B), $a=4$, $b=4$, $\eps=0.01$ and $I=0$, for various values of the coupling strength $J$ and noise level $\sigma$. (A) Individual trajectories for $20$ randomly chosen neurons in the network (dark blue: voltage, magenta: adaptation variable), together with the average voltage (light blue) and adaptation (red). The system shows perfect collective synchrony at intermediate values of $J$ and $\sigma$ (here, $J=1.5$ and $\sigma=1.5$, center); small coupling (bottom, $J=0.5$) or large noise (right, $\sigma=3$) leads to asynchrony, high coupling (top, $J=3$) or low noise (left, $\sigma=0.5$) to clamping. (B) Distribution of the voltage variable in each regime and at different times: clamped regimes show a tight unimodal stationary distribution (no dependence in time), asynchronous regimes either show a broad (large noise, right) or a bimodal (low coupling, right) stationary distribution. In the noise-induced regime, the distribution is periodic; starting e.g. around the resting state ($T=250$, blue), it progressively widens and moves  towards the excited regime, developing a bimodal shape as pioneers accumulate (yellow, $T=290$), until the whole distribution moves to the spiking region (purple, $T=300$) and comes back to rest (green, $T=320$). (C) Maximal amplitude (top) of the Fourier transform of the average adaptation and associated period (bottom) highlighting a wide region of synchronized oscillations.}
	\label{fig:phenomenon}
\end{figure*}

Coupling and noise have opposite effects on the collective dynamics: the former promotes coherence by pulling the voltage of each cell towards the average voltage the network generates, while noise reduces coherence by inducing random independent fluctuations of the voltage of each cell. Two regimes therefore arise for extremal values of coupling and noise (see Fig.~\ref{fig:phenomenon} and Supplementary Movie~M.1):
\begin{itemize}
	\item \emph{Asynchrony}: for coupling sufficiently low relative to a fixed noise level, interactions become too weak to induce macroscopically organized dynamics, and neurons fire at random times as they cross the separatrix (Fig~\ref{fig:phenomenon}-A, bottom). As a result, asynchronous trajectories emerge, and the system reaches a stationary distribution. That distribution displays a bimodal shape, with a majority of neurons around the resting potential, and a macroscopic fraction in the spiking regime or in the course of firing (Fig.~\ref{fig:phenomenon}-B, bottom)~\footnote{This distribution  is comparable to the distribution of voltage in the uncoupled case, and asymptotically converges to it for vanishing interaction~\cite{mischler2016kinetic}.}.
	
 Similarly, when noise is sufficiently large relative to a fixed coupling strength~\footnote{Noise and coupling indeed have opposite effects. Time rescaling $t'=\sigma^2 t$ indeed leads to a diffusion equation with unit noise variance, leading to an effective coupling coefficient $J/\sigma^2$. This rescaling also has an impact on other parameters in the system, including excitability and timescales, so there is no invariance in the system and one cannot assume that the impact of $J$ and $\sigma$ only depend on the ratio $J/\sigma^2$.}, neurons will fire asynchronously (fig~\ref{fig:phenomenon}-A, right): intrinsic noisy fluctuations in those regimes overwhelm the dynamics and coupling terms, allowing neurons to spike independently of the state of other neurons. A broad stationary distribution ensues, covering both resting and excited parts of the phase plane, which has essentially a unimodal shape skewed towards the spiking region (Fig.~\ref{fig:phenomenon}-B, right). 
 
 Common to both asynchronous regimes, the distribution of neuron variables reaches a stationary state covering rest and excited regimes; the absence of rhythmic activity is visible in the low maximal amplitude of the Fourier transform of the average voltage or adaptation variables (Fig.~\ref{fig:phenomenon}-C);
	\item \emph{Clamping}: For coupling sufficiently large (relative to a given noise level), or noise sufficiently small (relative to a coupling level), transitions to the excited regime are very rare, and the distribution of neurons remains clamped around the resting state (Fig.~\ref{fig:phenomenon}-A, top and left). Heuristically, for coupling large, the interaction term becomes prominent compared to the intrinsic dynamics and dominates noisy fluctuations; this term forces each neuron to remain in the vicinity of the empirical average of the voltage, preventing noise from leading neurons into individual excursions~\footnote{In the extreme case of coupling diverging to infinity, it can be shown that the distribution of state variables converges to a Dirac mass centered at a time-varying point whose dynamics follow the FitzHugh-Nagumo equation~\cite{quininao-touboul:18}, thus, after some time, accumulating at the resting state. Moreover, in that study, a Gaussian profile is predicted for large (but not infinite) $J$, opening the way in that regime to apply Gaussian moment closure as in~\cite{zaks}.}. For low noise, it is the rarity of transitions to the excited state that leaves the system passively ``clamped'' in the vicinity of the resting state.

	Common to both clamped regimes, the empirical distribution of neurons concentrates at a stationary solution centered at the resting state ((Fig.~\ref{fig:phenomenon}-B, left and top), and the low amplitude of the Fourier transform of the average voltage or adaptation variables underlines the absence of rhythmic behavior (Fig.~\ref{fig:phenomenon}-C). 
\end{itemize}

While the above-described regimes can be readily understood heuristically, the transition between these two regimes is much more surprising, and is associated with:
\begin{itemize}
	\item \emph{Noise-induced synchronization}: For intermediate values of coupling and noise, the dynamics are no longer stationary: the trajectories display sharp, perfectly periodic and synchronized macroscopic oscillations formed by all neurons firing synchronously within a small time interval (Fig.~\ref{fig:phenomenon}-A, center). In this regime, high amplitude oscillations of the distribution arise, as evidenced by the very peaked Fourier transform typical of periodic signals (Fig.~\ref{fig:phenomenon}-C). This non-stationarity is also visible in the distribution of the voltage at various times (Fig.~\ref{fig:phenomenon}-B, center). 
	
	The evolution of the distribution highlights the statistical physics of the phenomenon that will be described in more detail in the following sections. Consider for instance an initial state centered in the vicinity of the resting state (blue curve in Fig.~\ref{fig:phenomenon}-B, center). For noise sufficiently large and coupling sufficiently low, single neurons can overcome their attraction to the resting state, cross the separatrix and reach the excited state (neurons performing these transitions will be called \emph{pioneers} thereafter). These transitions do not lead to full spikes but partial deflections of the voltages, that can either lead to a spike or a return to rest. Yet, as more neurons perform such transitions,  the voltage distribution progressively broadens and develops a peak in the excited region, gradually shifting the voltage of resting neurons closer to the separatrix, making transitions to pioneers more likely and thus the peak of the voltage distribution in the excited regime more prominent (yellow curve in Fig.~\ref{fig:phenomenon}-B, center)~\footnote{Note that this significant bimodality precludes generalizing the Gaussian approximation of~\cite{zaks} to non-small noise levels; in particular, using the moment ODE for such parameters yields very distinct behaviors with no oscillation, arguing for the importance of bimodality in the model.}. This progressive buildup is followed by a very rapid transition where all neurons eventually switch to the excited state, and, synchronized with the other neurons, perform a full collective spike (a \emph{macroscopic spike}), associated with a very fast transition of the distribution to a unimodal one centered at the excited state. From that state, the distribution progressively returns to the vicinity of the resting state, where the process starts afresh. 
	
\end{itemize}

The regime of noise-induced synchronization is not a singular transition between clamping and asynchrony: a relatively wide, eye-shaped region in the plane $(J,\sigma)$ corresponds to such oscillations (Fig.~\ref{fig:phenomenon}-C), arising through a sharp transition from clamping (high-amplitude strongly periodic oscillations with very low frequency arise from clamping regimes), and smoothly transitioning to asynchrony. This suggests a transition via a Hopf bifurcation on the high noise side, and a homoclinic bifurcation on the low noise side, that will be shown to be a general observation in systems of coupled excitable elements in section~\ref{sec:Universality}.

\section{Statistical Mechanics of noise-induced synchronization}\label{sec:NoiseInducedModel}
Noise-induced synchronization is thus associated with two remarkable phenomena: the buildup of a bimodal voltage distribution with spontaneous variations of the relative amplitude of the two peaks, as well as a sharp and sudden transition of all neurons to the excited regime where a macroscopic spike is emitted. 

In section~\ref{sec:chainReaction}, we show that in the absence of noise, there exists a critical proportion of pioneers $\alpha=\alpha_c$ below which pioneers return to rest without firing a spike, and above which a chain reaction is triggered inducing a macroscopic spike. This leads us to conjecture that noise-induced oscillations arise when stochastic dynamics naturally lead the system to exceed this critical fraction. We thus study in section~\ref{sec:Transitions} the transitions between resting and pioneer states, allowing to account for the dynamics of the bimodal distribution and to delineate regimes where the network spontaneously reaches the critical fraction of pioneers needed to synchronize (thus, enters the noise-induced oscillations regime), or remains below that threshold (leading to stationary solutions).

\subsection{Nonlinear Dynamics of Macroscopic Spikes}\label{sec:chainReaction}
Macroscopic spikes stand out as sudden, sharp and dramatic events affecting all neurons, and arising as the fraction of neurons in the excited state progressively increases. We show here that this sudden switch can be associated with a nonlinear change of the stability of the resting state as the number of pioneers increases, independently of the stochastic fluctuations.

\begin{figure*}[ht]
\includegraphics[width=15cm,height=15cm,keepaspectratio]{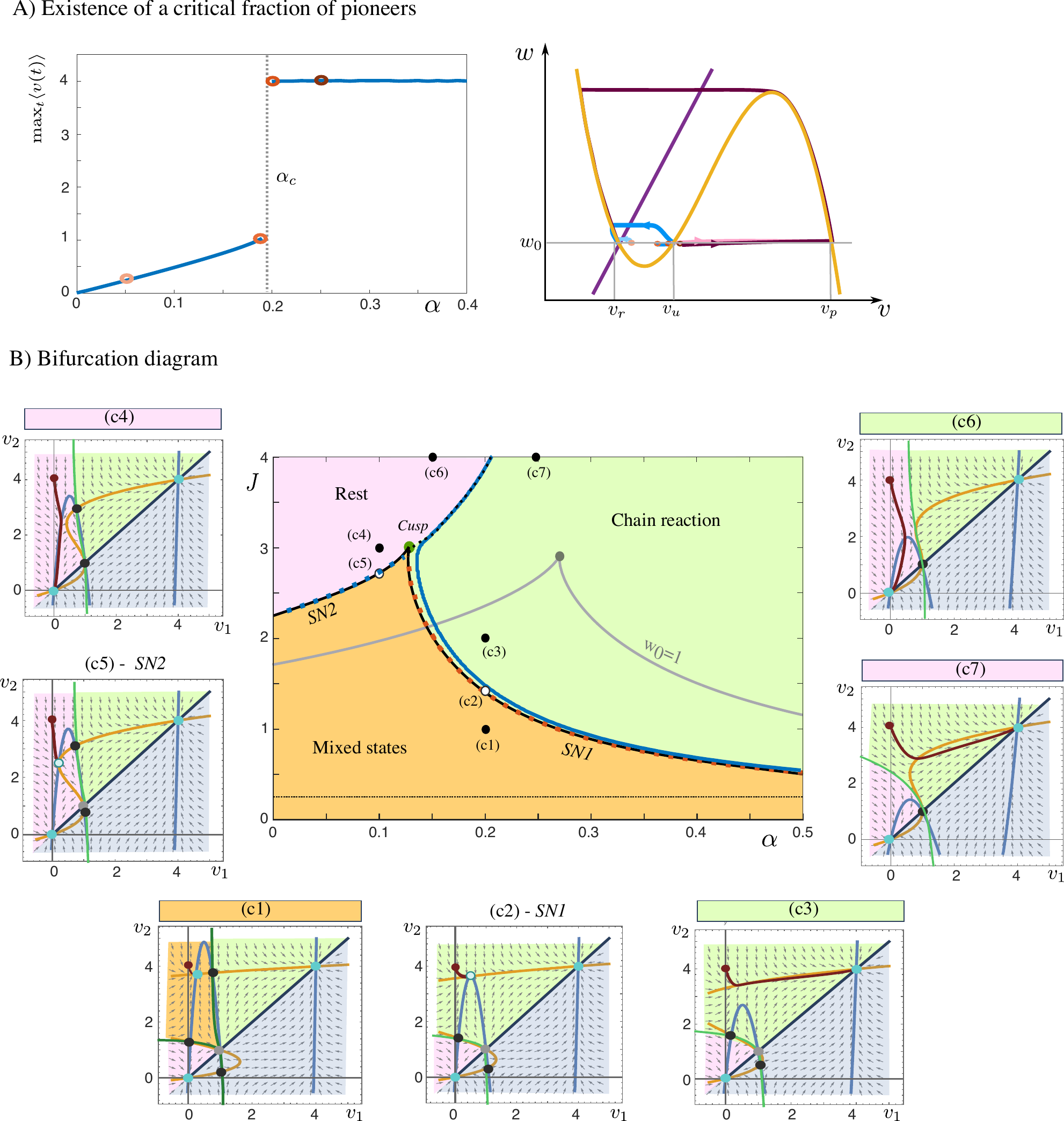}
\caption{Chain Reaction. (A) left: Maximal value of the average voltage for network~\eqref{eq:FhN} in the absence of noise ($\sigma=0$) as a function of the initial proportion of pioneers $\alpha$ ($n=1\,000$, $a=4$, $\eps=0.01$, $b=4$, $J=1.5$, $I=0$) shows a sudden discontinuity at a critical value $\alpha_c$, where (right) trajectories in the phase plane (yellow: $v$-nullcline, purple: $w$-nullcline) switch from rapid returns to rest for $\alpha<\alpha_c$ (light blue, $\alpha=0.05$, blue: $\alpha=0.19$, lighter two shades of red circles in the right diagram) to long collective spiking excursions for $\alpha>\alpha_c$ (pink, $\alpha=0.21$, red, $\alpha=0.25$, darker two shades of red circles in the right panel). Initial condition (averaged) is depicted by a colored circle. Gray line: $w_0$, depends on the level of  noise; the values $v_r$, $v_u$ and $v_p$ are intersections of the line $w=w_0$ and the cubic nullcline. (B) Critical fraction of pioneers associated with the chain reaction as a function of coupling $J$ in the two-dimensional system~\eqref{eq:FhN} (blue curve) or in the simplified equation~\eqref{eq:Alpha} (dashed blue: transition from rest to mixed equilibria, dashed red: transition from mixed equilibria to pioneers, dotted black: transition from rest to pioneer), together with the bifurcations of the simplified one-dimensional model~\eqref{eq:Alpha} with $w_0=0$ (black) or $w_0=1$ (gray), showing two saddle-node bifurcations curves (SN1 \& SN2) joined at a cusp bifurcation. (c1-c5) Typical phase portraits of the simplified system~\eqref{eq:Alpha}. Dark blue: identity line (irrelevant phase space grayed), Yellow and blue curves: $v_1$ and $v_2$ nullclines, cyan circles: stable fixed points, gray circles: unstable fixed points, black circles: saddles. Green lines represent the stable manifold of the saddles, and partition the phase-space into the attraction basin of stable fixed points (green: chain reaction, orange: mixed equilibrium, pink: rest, as in (c)). Arrows represent the direction of the flow. The dark red curve represents the trajectory starting from $v_1=v_r$ and $v_2=v_p$. (c1) $\alpha=0.2$ and $J=1$, (c2) $\alpha=0.2$ and $J= 1.4$ near SN1, (c3) $\alpha=0.2$ and $J=2$, (c4) $\alpha=0.1$ and $J=3$, (c5): $\alpha=0.1$ and $J=2.73$ near SN2, (c6) $J=4$ and $\alpha=0.15$, (c7) $J=4$ and $\alpha=0.25$.}
\label{fig:ChainReaction}
\end{figure*}

To this purpose, we investigate the behavior of a set of $n$ neurons satisfying equation~\eqref{eq:FhN} in the absence of noise as the initial fraction of neurons in the pioneer state, $\alpha$, is varied. For $\alpha$ sufficiently small, we indeed observed no macroscopic spike generated (Fig~\ref{fig:ChainReaction}): the maximal value of the average voltage remains bounded below the spike level (Fig~\ref{fig:ChainReaction}-A), and trajectories of the system show a direct return of all neurons to rest (Fig~\ref{fig:ChainReaction}B). However, a sudden and sharp transition occurs at a critical fraction of pioneers $\alpha_c$: the presence of a sufficient number of pioneers produces a strong attraction towards the excited state, leading to a \emph{chain reaction} during which all remaining resting neurons transition to pioneers, followed by a macroscopic spike. For these initial levels of pioneers, the maximal value of the average voltage jumps to the spiking level (Fig~\ref{fig:ChainReaction}-A) and the average trajectory in the phase plane (Fig.~\ref{fig:ChainReaction}B) displays a spike and return to rest.

To determine analytically $\alpha_c$ and quantify precisely its dependence upon $J$, we consider the nonlinear evolution of the average voltage of resting ($v_1$) and pioneer ($v_2$) neurons assuming that the transition occurs in a timescale faster than the adaptation variable (i.e., adaptation considered fixed equal to some value $w_0$), a relevant approximation here owing to the slow evolution of the adaptation variable during the transition~\footnote{Note also that neglecting the evolution of the adaptation variable has little impact on the collective dynamics since electrical coupling only relies on voltage.}. For a given proportion of pioneers $\alpha$, the evolution of $v_1$ and $v_2$ is thus approximated by the two-dimensional ODE:
\begin{equation}\label{eq:Alpha}
\begin{cases}
v_1' = f(v_1)-w_0+J\alpha(v_2-v_1) & v_1(0)=v_r\\
v_2' = f(v_2)-w_0-J(1-\alpha)(v_2-v_1) & v_2(0)=v_p.
\end{cases}
\end{equation}
where $w_0$ corresponds to an adaptation value such that the dynamics of a single particle (uncoupled system):
\begin{equation}\label{eq:uncoupled}
	v' = f(v)-w_0
\end{equation}
 display three equilibria: a resting state $v_r$ on the leftmost branch of the cubic function $f$, a pioneer state $v_p$ on the rightmost branch (values appearing as initial conditions in eq.~\ref{eq:Alpha}) and an unstable fixed point $v_u$ on the middle branch acting as a separatrix between the attraction basins of the two equilibria~\footnote{If a single equilibrium exists, then the system cannot support a mixture of pioneers and resting neurons, and no critical value of $\alpha$ exists, as, regardless of the initial proportion of pioneers, all neurons eventually reach the same equilibrium; if that equilibrium is on the pioneers branch, a spike will be fired regardless of $\alpha$, while if the equilibrium is on the resting state, no chain reaction occurs.}. We are interested in solutions of these equations with $v_1\leq v_2$. 
 
The dynamical system~\eqref{eq:Alpha} features at least three fixed points with identical voltage $v_1=v_2 \in \{v_r,v_u,v_p\}$  (Fig.~\ref{fig:ChainReaction}B). These fixed points have the same stability as the associated voltage in 1-dimensional uncoupled FitzHugh-Nagumo system~\eqref{eq:uncoupled}. Indeed, the Jacobian matrix of the system~\eqref{eq:Alpha} at an arbitrary point $(v_{1}^*,v_{2}^*)$ reads:
\[
\left(
\begin{array}{cc}
f'(v_1^*)-J\alpha & J\alpha\\
J(1-\alpha) & f'(v_2^*)-J(1-\alpha).
\end{array}
\right)
\]
For $v_1^{*}=v_2^{*}=v_r$ or $v_p$ (the two stable fixed points of~\eqref{eq:uncoupled}), the trace of the Jacobian matrix is strictly negative and its determinant, $f'(v^*)^2-J f'(v^*)$, is strictly positive (since $f'(v^{*})<0$ for a stable fixed point), implying stability of the fixed points $(v_r,v_r)$ and $(v_p,v_p)$ for the two-dimensional system. If $v_1^*=v_2^*=v_u$ the unstable fixed point of~\eqref{eq:uncoupled}, the trace reads $2f'(v_u)-J$ and the determinant $f'(v^{*})(f'(v^{*})-J)$. The determinant is non-negative when $f'(v^{*})>J$, in which case the trace is positive. The fixed point $(v_u,v_u)$ is thus always unstable for system~\eqref{eq:Alpha}: it is a saddle when $f'(v^{*})<J$ and an unstable node for $f'(v^{*})>J$. 

The equilibria with identical voltage correspond to a synchronization of the network: $(v_r,v_r)$ corresponds to all pioneers returning to rest, and $(v_p,v_p)$ to all resting neurons reaching the excited state. In addition to these attractors, mixed equilibria (i.e., with $v_1<v_2$) may also exist, when the system is able to support a mixture of a fraction $\alpha$ of pioneers and $(1-\alpha)$ of resting neurons (e.g., in the asynchrony regime). A chain reaction occurs if, starting from the initial condition $(v_1(0)=v_r,\;v_2(0)=v_p)$, the system reaches the fixed point $(v_p,v_p)$, or in other words $(v_r,v_p)$ belongs to the attraction basin of $(v_p,v_p)$. We thus analyzed the geometry of the phase plane of system~\eqref{eq:Alpha}, and found that, as parameters are varied, the initial condition $(v_r,v_p)$ could suddenly switch attraction basins, associated either with smooth changes in the shape of these basins, or organized by the stable manifold of a saddle mixed equilibrium (Fig.~\ref{fig:ChainReaction}C).

To characterize these transitions, we first computed the two-parameter bifurcation diagram of equation~\eqref{eq:Alpha} as a function of the coupling strength $J$ and the proportion of pioneers $\alpha$. This diagram displays two branches of saddle-node bifurcations (black curves) merging at a cusp bifurcation. These saddle-node curves delineate the region where mixed equilibria exist. We found that these bifurcations coincide exactly with transitions in the eventual state of the system starting from the initial condition $(v_r,v_p)$ (dashed blue and red curves, obtained by extensive simulations of the system). Moreover, we recover precisely the value $\alpha_c$ corresponding to the transition found for the full system~\eqref{eq:FhN} at $J=1.5$ and with $\sigma=0$ (Fig.~\ref{fig:ChainReaction}A). More generally, when $J$ is varied, we found an excellent agreement between the critical fraction $\alpha_c$ and the evaluations obtained with the simplified system~\eqref{eq:Alpha} (blue curve in Fig.~\ref{fig:ChainReaction}).

The coincidence between saddle-node bifurcations and switches in the attractor to which solutions of system~\eqref{eq:Alpha} converges can be inferred analyzing the geometry of vector field of equation~\eqref{eq:Alpha}. Attraction basins in this two-dimensional system are governed by the stable and unstable manifolds of the saddle fixed points, depicted in green in Fig.~\ref{fig:ChainReaction}c1-c7. We observe that whether or not the initial condition belongs to the attraction basin of rest or pioneer strongly depends on the presence of mixed saddles (rather than the specific shape of those manifolds), accounting for the perfect agreement between saddle-node bifurcations and switches in the fixed point towards which the system converges. Mixed equilibria do not exist for $J$ larger than the value associated with the cusp. For such parameters, $(v_u,v_u)$ is a saddle, and its stable manifold splits the phase space into the attraction basins of $(v_r,v_r)$ and $(v_p,v_p)$. Smooth changes in the shape of that manifold lead to a switch between resting and pioneer, and the associated value of $\alpha$ associated with the transition is represented as a dotted line in Fig.~\ref{fig:ChainReaction} and closely agrees with the chain reaction separatrix (blue curve) computed for the full system. 

Clamping at high coupling strength may be associated with the incapacity of the system to support a mixture of pioneer and resting neurons for $J$ larger than the value associated with a cusp (the full system returning to rest either directly of after one spike depending on the initial $\alpha$). In sharp contrast, coupling values $J$ small enough (below the dashed line in Fig.~\ref{fig:ChainReaction}C) allow the existence of mixed equilibria for any initial proportion of pioneers $\alpha$: coupling is not sufficient to promote a collectively coherent behavior, leading to asynchronous behaviors.

\begin{figure}[h]
	
	\includegraphics[width=\columnwidth]{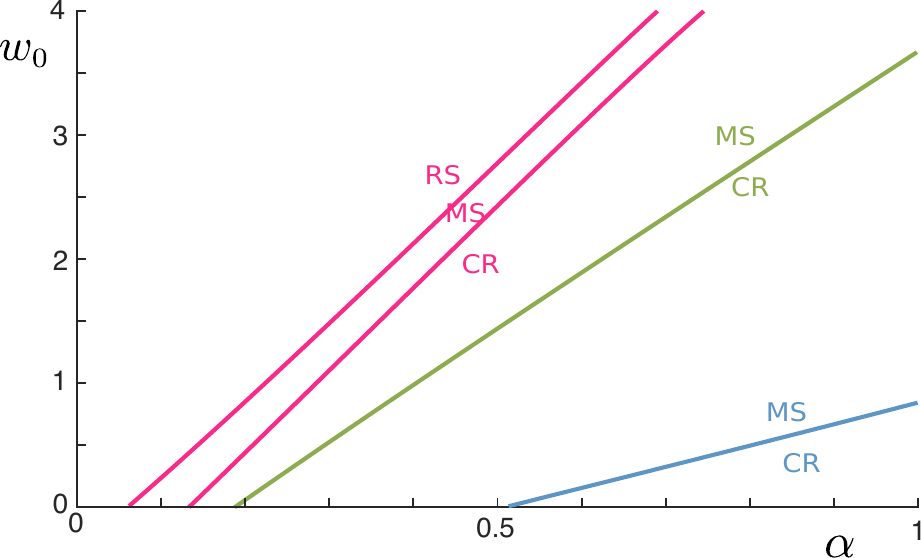}
	\caption{Two-parameter bifurcation diagram of the simplified system~\eqref{eq:Alpha} with respect to $\alpha$ and $w_0$ for various values of the coupling strength: $J=2.5$ (pink), $J=1.5$ (green) and $J=0.5$ (blue); RS: attraction to the resting state, MS: attraction to mixed states, CR: chain reaction: attraction to the pioneer fixed point. No codimension-two bifurcation is found, indicating no qualitative change in the dynamics as $w_0$ is modified (e.g., when noise is varied). }
	\label{fig:Codim2Chain}
	\end{figure}

The bifurcations and transitions found depend on the value of $w_0$, which, in the original model with noise~\eqref{eq:FhN}, depends both on $\sigma$ and $J$. To appreciate how $w_0$ alters the transitions found in the system, we computed the two-parameter bifurcation diagram of system~\eqref{eq:Alpha} as a function of $\alpha$ and $J$ for a distinct value of $w_0$ (Fig.\ref{fig:ChainReaction}C, gray curve, $w_0=1$), and found a qualitatively similar bifurcation diagram as originally obtained for $w_0=0$, but essentially shifted to larger values of $\alpha$. Heuristically, a larger $w_0$ would correspond to larger noise (or weaker coupling), and the system would necessitate a larger fraction of pioneers to display the chain reaction. The absence of qualitative dependence in $w_0$ was confirmed in Fig~\ref{fig:Codim2Chain} representing the two-parameter bifurcation diagram of system~\eqref{eq:Alpha} as a function of $\alpha$ and $w_0$ for distinct values of $J$. We found monotonically increasing bifurcation lines and no codimension-two bifurcation, showing that the qualitative features outlined above persist for various values of $w_0$. 

We thus conclude that a chain reaction may arise when the population is composed of a sufficiently large fraction of pioneers. We now relax our assumption of considering fixed values of $\alpha$, and turn our attention to the stochastic transitions between resting and pioneer that govern the evolution of $\alpha$.

\subsection{Reaching the critical proportion of pioneers}\label{sec:Transitions}

The question that arises is thus whether the critical fraction of pioneers $\alpha_c$ is reached spontaneously by the stochastic system. In the presence of noise, neurons at rest may indeed transitions to pioneer, and reciprocally~\eqref{eq:FhN}; the fraction of pioneers will thus vary in time according to the rates of transitions. To describe this evolution, we now quantify the rates at which resting neurons transition to pioneers and reciprocally. 

\subsubsection{Stochastic Transitions between pioneer and resting states}\label{sec:purelyStochastic}
Characterizing the rate of transition of a stochastic particle in a multi-well potential is a classical and widely studied question in the domain of stochastic analysis~\cite{freidlin-wentzell:98,kramer}. Most results are derived in the small noise limit, and for Hamiltonian systems. The problem of noise-induced synchronization we are studying here challenges classical theory in many ways. In particular, noise-induced oscillations arise for non-vanishing noise, the dynamics are not Hamiltonian, noise modifies the dynamics (potential and equilibria in a Hamiltonian analogy) and the transitions are collective, i.e. transition rates depend on the positions of all other particles through the coupling term. In supplementary section~\ref{sec:Kramer}, we discuss in more detail these questions, highlight the difficulty to define a double-well potential for the system and how that putative potential depends on noise and on the distribution of neuron voltages (chiefly through $\alpha$), and how theoretical estimates may deviate from the effective rates of transitions.


To numerically evaluate the transition rates from rest to pioneer and reciprocally, we again reduced the system to a one-dimensional equation on the voltage only, assuming a fixed value of the adaptation variable $w=w_0$ during the transition phase. Under this hypothesis, and between two transitions, each neuron satisfies the equation:
\begin{equation}\label{eq:Particle}
dv_t=\Big(f(v_t) -w_0 + J ((1-\alpha) v_r+\alpha v_p-v_t)\Big)\, dt + \sigma dW_t.	
\end{equation}
where $\alpha$ is the current fraction of pioneers (fixed between two consecutive transitions), and $v_p$ and $v_r$ are the pioneer and resting voltage, approximated as the largest and smallest solution of 
\[f(v)=w_0.\]

\begin{figure*}
	\begin{center}
		\includegraphics[width=.6\textwidth]{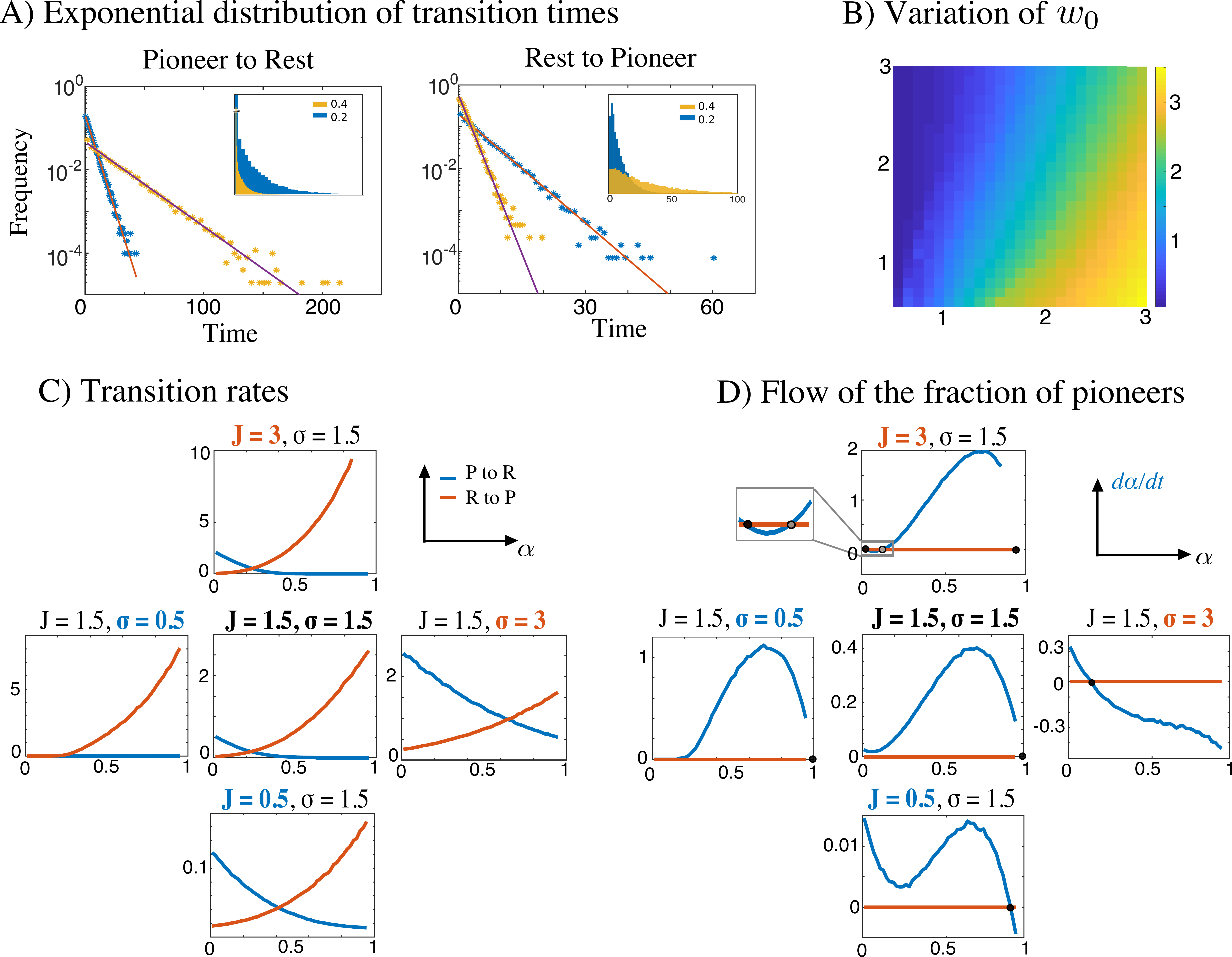}
	\end{center}
	\caption{Stochastic transitions between resting and pioneer states: (A) Histogram of the transition time from pioneer to rest (left) and rest to pioneer (right), in semi-logarithmic scale (inset: linear scale) for $\alpha=0.2$ (blue) or $\alpha=0.4$ (yellow). The corresponding maximum likelihood fit with an exponential distribution shows an excellent match (solid lines, red: $\alpha=0.2$, purple: $\alpha=0.4$). Parameters as in Fig.~\ref{fig:phenomenon} with $J=1.5$, $\sigma=1.5$ and $\alpha=0.2$; a sample $15\,000$ independent simulations of equation~\ref{eq:Particle} was used. (B) Variation of $w_0$ as a function of $\sigma$ and $J$, computed as the median value of adaptation at the transition for the original system with $n=4\,000$ neurons. (C) Transition rates obtained from the maximum likelihood estimator with the exponential distribution as a function of $\alpha$ (blue: Pioneer to rest, red: rest to pioneer, 15\,000 simulations of~\eqref{eq:Particle} for the 5 situations considered in Fig~\ref{fig:phenomenon}-A). (D) Flow of the fraction of pioneers (righthand side of eq.~\ref{eq:alpha_dyn}, blue curve vs $0$ in red). } 
	\label{fig:Transitions}
\end{figure*}

The first step of this program thus consists in evaluating $w_0$ numerically as a function of those parameters. To this end, we simulated the full system eq.~\ref{eq:FhN} and computing the median adaptation at transitions from rest to pioneer (see Fig.~\ref{fig:Transitions}(B))~\footnote{Quantitative estimates of $w_0$ as a function of those parameters is a complex task, owing to the fact that relevant noise levels are non-vanishing and to the collective nature of the stabilization. These aspects prevent from readily extending methods used in the small noise limit for single particles. For a single particle and in the scaling regime of SISR, Deville and collaborators~\cite{deville2005two} estimated the dependence of $w_0$ in $\sigma$. In detail, for a single neuron with small noise and adaptation timescale with a specific scaling ($\sigma^2\log(\eps^{-1})$ of order 1), an analytic approach allows quantifying $w_0$ as a function of $\sigma$ using stochastic estimates combining the relative values of exponential attraction towards the equilibrium invariant manifold (leftmost or rightmost branch of the cubic nullcline) and the exponential switching rate across the separatrix obtained by Freidlin-Wentzell estimates. These methods fail when noise is non-vanishing, as large-deviations estimates are asymptotic results for vanishing noise.}. Using this estimate, we next systematically evaluated, for various pairs of $(J,\sigma)$, the distribution of transition times for a particle satisfying eq~\ref{eq:Particle} for multiple (typically, $15\,000$) realizations of the process with independent noise and independent random initial condition within the pioneer or resting regime of the uncoupled system~\footnote{We chose normally distributed initial conditions with a truncation. For pioneers to rest (resp. rest to pioneer) transition, this initial condition was centered at $v_p$ (resp. $v_r$) and truncated to ensure $v> v_u$ (resp. $v<v_u$), where $(v_r,v_u,v_p)$ are defined as the solutions of $f(v)=w_0$.}.

A particle is considered to switch attractors if, starting on one side of the separatrix (embodied by the unstable fixed point $v_u$), it ends up dwelling on the other side. To avoid considering transient passages of a particle through the separatrix not corresponding to actual transitions, we used a confidence interval to determine transition times: a pioneer (resp., a rest) neuron was considered to have switched to rest (resp., pioneer) if its voltage exceeds $\gamma v_r +(1-\gamma) v_u$ (resp., goes below $\gamma v_p +(1-\gamma) v_u$) with $\gamma=0.1 $\footnote{The choice of the threshold did not affect qualitatively the results, yet a threshold exactly at $v_u$ was prone to detecting false transitions, and our choice of threshold has the advantage of taking into account the distance between the separatrix and the equilibrium compared to fixed threshold that would under-estimate transitions towards the equilibrium closest from the separatrix. }. We observed that the transition times are exponentially distributed with a rate that depends upon the proportion of pioneers (see Fig.~\ref{fig:Transitions} (A)). We confirmed that the data is indeed statistically consistent with an exponential distribution fitting a rate using the maximum likelihood estimator and a Kolmogorov-Smirnov test. For $\alpha=0.2$ (Fig.~\ref{fig:Transitions}-A), we found a rate of transition from pioneer to rest (resp., rest to pioneer) equal to $\lambda=0.20$ ($\lambda=0.17$), with a goodness of fit (Kolmogorov-Smirnov) $0.013$ (resp. $0.0153$) and a p-value $p<0.015$ (resp., $p=0.017$, see~\cite{clauset2009power} for details on the statistical test and code), confirming that the data is perfectly consistent with an exponential.

Finding an exponential distribution opens the way to an important simplification of the system. Indeed, because exponential transitions are memoryless, the evolution of the proportion of pioneers can now be modeled as a continuous-time finite-state Markov process. In detail, assuming that the distinct neurons have independent and identically distributed transition times (Boltzmann's standard molecular chaos hypothesis in large interacting particle systems, or propagation of chaos, proved for the FitzHugh-Nagumo network in~\cite{mischler2016kinetic}) the number of pioneers $(P_t)_{t\geq 0}$ forms a birth-and-death Markov process on the finite state space $\{0,\cdots,n\}$ with transitions:
\[\begin{cases}
x \to x+1 \text{\;\;with rate\;\;} (n-x) \, K_{RP}(\frac x n) & \text{for any } x<n,\\
x \to x-1 \text{\;\;with rate\;\;} x \, K_{PR}(\frac x n) & \text{for any } x>0.
\end{cases}
\]
where $K_{RP}$ and $K_{PR}$ are the transition rates from rest to pioneer and reciprocally. Therefore, for $n$ large, using the Kolmogorov equation for that Markov chain, we find that the proportion of pioneers at time $t$, $\alpha(t)=P_t/n$, satisfies the ordinary differential equation:
\begin{equation}\label{eq:alpha_dyn}
	\frac{d\alpha}{dt} = (1-\alpha)\, K_{RP}(\alpha) - \alpha K_{PR}(\alpha).
\end{equation}

The fixed points of that one-dimensional equation, representing the steady-state proportions of pioneers, are given by the implicit equation:
\[\alpha^\star=\frac{K_{RP}(\alpha^\star)}{K_{RP}(\alpha^\star)+K_{PR}(\alpha^\star)},\]
and these are stable equilibria when:
\[(1-\alpha^{\star})K_{RP}'(\alpha^{\star}) -\alpha^{\star} K_{PR}'(\alpha^{\star}) - (K_{RP}(\alpha^{\star})+K_{PR}(\alpha^{\star}))<0.\]

To determine equilibrium proportions of pioneers and their stability as a function of the parameters, we systematically evaluated the distribution of transition times for various values of $\alpha$ and fitted an exponential distribution using the maximum likelihood estimator (15\,000 samples for each condition). We observed, as expected, that the rate of transition from rest to pioneer increases with $\alpha$ while the rate of the reciprocal transition decreases (see Fig.~\ref{fig:Transitions}(C)). Indeed, the larger $\alpha$, the faster resting neurons reach the pioneer state, while pioneer states are stabilized for $\alpha$ large and transitions to rest rarer. From these transition curves, we computed the flow of $\alpha$ (equation~\eqref{eq:Alpha}) as a function of coupling $J$, noise $\sigma$, and proportion of pioneers $\alpha$. We observe clear transitions in the shape of the flow, associated with the transitions from clamping to synchrony and then asynchrony, as visible in Fig.~\ref{fig:Transitions}(D) for the values of $J$ and $\sigma$ highlighted in Fig.~\ref{fig:phenomenon} (see also Fig.~\ref{fig:CorrectedAlpha}B for a more exhaustive representation of the flow, corrected by the spontaneous spike rate of pioneers introduced in section~\ref{sec:withSpikes}). 

Although our numerical estimations neglect the transitions from pioneers to rest subsequent to a spike (introduced in the next section), it already accounts for three of the five regimes observed in Fig.~\ref{fig:phenomenon}:
\begin{itemize}
	\item In the synchronized regime (central regime in Fig.~\ref{fig:phenomenon}(A)), the flow is uniformly positive, indicating that $\alpha$ is strictly increasing regardless of the initial condition. Therefore, neurons accumulate rapidly within the pioneer region, reach the critical proportion $\alpha_c$, undergo the chain reaction through a macroscopic spike and return to $\alpha=0$ where the process starts afresh. Because the typical time for one given transition is significantly smaller than the time of a spike (see section~\ref{sec:withSpikes}), the model accurately accounts for the observations on the full system. Moreover, we note that the time for the proportion of pioneers to reach $\alpha_c$ is deterministic (it is the crossing time of $\alpha_c$ of the solution of equation~\eqref{eq:alpha_dyn} starting from $\alpha=0$), accounting for the regularity of the periodic behavior observed in Fig.~\ref{fig:phenomenon}. 
	\item In the clamped regime for large $J$ (top, Fig.~\ref{fig:Transitions}(D)), we observe that the fraction of pioneers shows bistable dynamics, with a stable fixed point at $\alpha=0$ (clamped state), a stable fixed point at $\alpha=1$ (chain reaction regime), and a third unstable fixed point separating the attraction basins of both equilibria near the chain reaction threshold (for large $J$, $w_0$ is close from $0$ and the chain reaction threshold is slightly below $0.2$, see Fig.~\ref{fig:ChainReaction}D). 

	Depending on the initial $\alpha$, either all neurons return to rest, or all neurons reach the pioneer state, fire a macroscopic spike and return to the clamped regime near equilibrium. The computed curves seem to indicate that this transition occurs through a saddle-node bifurcation in the rate equation~\eqref{eq:Alpha}, as the quadratic behavior of the flow for small $\alpha$ progressively shifts up and becomes tangent to the $0$ line. As the system approaches the transition, the time taken by the system to trigger a spike decays to zero, accounting for the notable slowing down of the oscillations near this transition (see Fig.~\ref{fig:phenomenon}). 
	\item Eventually, for large $\sigma$, the transition rate from pioneer to rest and back again shows a less sensitive dependence in $\alpha$: noise becomes sufficient to induce transitions in both directions for any value of $\alpha$, and starts dominating the interaction terms. The fraction of pioneers $\alpha$ thus rapidly converges to the unique stable fixed point of the associated system, close from $0.2$ here. At this value of $\alpha$, the number of resting neurons transitioning to pioneer is balanced by the number of pioneers returning to rest, and a stationary asynchronous firing regime ensues. 
\end{itemize}

In all those three cases, the rates of transitions are relatively large compared to the duration of a spike, and therefore the accumulation of pioneers and transient variations of $\alpha$ occurs prior to any neuron firing a spike. This is not the case of low noise clamping or low coupling asynchronous regimes. In the low noise regime, both rates of transitions are very low. In the low-coupling case, rates of transitions are no more low, but the transition from rest to pioneers are compensated by the reciprocal transition arising at a similar rate. In both cases, the very slow evolution of the fraction of pioneers ensuing allows neurons that have transitioned early to pioneers to spike and return to rest before accumulation of neurons in the pioneer state. This phenomenon shall be crucial when the typical time for reaching critical value $\alpha_{c}$ is larger than the typical spike duration (or when it is infinite, i.e. when $\alpha_c$ is never reached).

\subsubsection{Modeling Returns to Rest after Spiking} \label{sec:withSpikes}

To account for  this phenomenon, we added a corrective term to equation~\eqref{eq:alpha_dyn} considering the rate at which pioneers return to rest after spiking. To this end, we computed the typical time for a pioneer to fire a spike. The spike is composed of three main phases: an upstroke where the voltage climbs up the rightmost branch of the stable part of the cubic nullcline from $w=w_0$ up to a value $w_1$ where typical trajectories leave that branch (much like $w_0$, the value of $w_1$ depends on noise and coupling), a rapid jump to the leftmost branch of the cubic at $w=w_1$, and then a downstroke where the trajectory decays along the right branch of the cubic down to $w=w_0$. Neglecting the rapid switching time between the upstroke and downstroke phase, we can express analytically the spike time as:
\begin{equation}\label{eq:T_spike}
	T_s=\frac 1 \eps \left(\int_{w_0}^{w_1}  \frac{dw}{b v_p(w) - w}+ \int_{w_1}^{w_0} \frac{dw}{b v_r(w) - w}\right)
\end{equation}
where $v_p$ and $v_r$, as defined above, are the right and left solutions of the cubic polynomial equation 
\[f(v)-w+I=0.\] 
The above formula highlights the fact that the spike time is of order $\eps^{-1}$, much longer compared to the stochastic fluctuations, but not necessarily longer than the time needed to accumulate neurons in the pioneer state. We thus used the above formula and the classical analytical expressions of $v_r$ and $v_p$ (we chose the trigonometric form for $v_r$ and $v_p$ due to Fran\c{c}ois Vi\`ete) and a numerical evaluation of $w_1$ (similar methodology as used for $w_0$) to compute the spiking time. 

We used this estimate of the typical spike duration, and the fact that $n$ is considered large, to model the fraction of pioneers returning to rest after a spike as a ceaseless leak of the density of pioneers to rest at a rate $T_s^{-1}$:
\begin{equation}\label{eq:alphawithSpikes}
	\frac{d\alpha}{dt} = (1-\alpha)\, K_{RP}(\alpha) - \alpha \,\left(K_{PR}(\alpha)+T_s^{-1}\right).
\end{equation}

\begin{figure*}
  \centering
    \includegraphics[width=.9\textwidth]{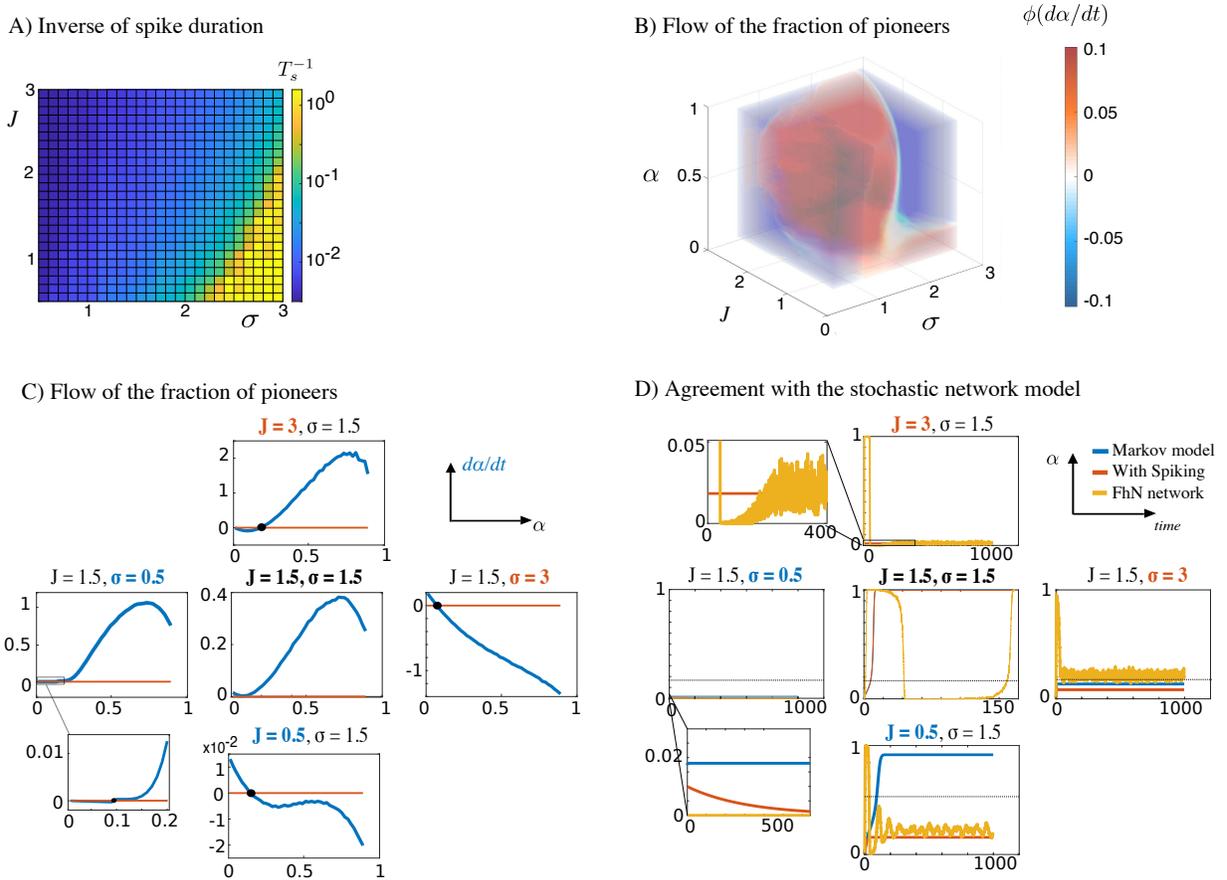}
  \caption{Stochastic and spiking transitions: combining stochastic transitions and spiking recovers qualitatively and quantitatively the phenomenology of the stochastic FitzHugh-Nagumo network of Fig.~\ref{fig:phenomenon}. (A) Numerical evaluation of the inverse spike duration (formula~\eqref{eq:T_spike}) as a function of $\sigma$ and $J$. (B) Flow of the fraction of pioneers (righthand side of equation~\eqref{eq:alphawithSpikes}) as a function of $\sigma$, $J$ and $\alpha$ (four-dimensional representation using pcolor3 matlab routine~\cite{Greene}). The system recovers the clamping fixed point with $0$ pioneers in the low noise or high coupling regimes, non-trivial fixed points for large noise or low coupling regimes, and a uniformly positive flow (chain reaction and noise-induced synchronization) for intermediate values (red sphere-like surface). For legibility, the flow was smoothed and thresholded through the function $\phi: x\mapsto (1+\tanh(4x))/2$. 
  (C) Flow of the fraction of pioneer $\alpha$ in the five situations considered in Fig.~\ref{fig:phenomenon}(A), recovers the appropriate dynamics in all cases qualitatively. (D) Quantitative agreement of the simplified model~\eqref{eq:alphawithSpikes} and the stochastic network model. Simulations of the number of pioneers in the original FitzHugh-Nagumo network (yellow) shows a good qualitative agreement with the one-dimensional stochastic transitions models with spiking (red), which significantly differs from the model without spiking (blue) in the low coupling case (predicting noise-induced oscillations) or low noise limit (the $\alpha$ remains almost constant due to very low rates, while it slowly returns to $0$ when spiking is considered). When it exists, the critical fraction of pioneers $\alpha_c$ associated with the chain reaction is depicted as a black dashed line. }
  \label{fig:CorrectedAlpha}
\end{figure*}

We confirmed that this correction is essentially negligible in the case of noise-induced oscillations, large noise asynchrony or large coupling clamping, but has important qualitative implications in the low noise or small coupling regimes, as visible in Fig.~\ref{fig:CorrectedAlpha}. Indeed, we observe that for small noise, very small values of $\alpha$ are stabilized, owing to the rarity of transitions are low $\sigma$. More generally, in the clamping regime the system stabilizes to a fraction of pioneers $\alpha$ at which transitions from rest to the excited state are compensated by stochastic and firing returns: small perturbations of this equilibrium towards smaller $\alpha$ will progressively be compensated by noise, and a slight deviation towards a higher $\alpha$ below the chain reaction will be damped as neurons fire a spike and return to rest. In this regime however, the dynamics are relatively sensitive, as the flow converges, when $\sigma\to 0$, to a discontinuous flow (with a discontinuity at the critical $\alpha=\alpha_c$ associated with the chain reaction). This sharpness is visible in our simulations (Fig.~\ref{fig:CorrectedAlpha}). 

When coupling is small, the corrected system taking into account the return of pioneers to rest through a spike also recovers the observed behavior of the emergence of a stable fixed point. In this regime, collective behaviors are less likely to arise as each cell is mostly driven by its own noise and excitable dynamics, essentially dominating the coupling term. Because of low coupling, while rates of transition remain of the same order of magnitude, the absence of a strong influence of network dynamics implies that the rates essentially compensate. We indeed observe that transition times are of the same order of magnitude of the duration of a spike (notice the maximal rate of transition on the order of $15.10^{-3}$ in Fig.~\ref{fig:Transitions}(C)). Therefore, adding the return to rest term associated with spikes has a highly non-trivial impact, and yields the emergence of a stable fixed point for $\alpha$ below the critical value, indicating the emergence of a stationary probability to be in the pioneer state. In other words, because neurons are asynchronous, the fraction of neurons in the pioneer or rest regime is the stationary mass of the distribution of a single neuron within those sets. 

We confirmed that the model precisely accounts for the evolution of the number of pioneers as a function of time, starting from $\alpha=0$. We depict in Fig.~\ref{fig:CorrectedAlpha}(D) the fraction of pioneers computed in the stochastic FhN network~\eqref{eq:FhN} together with the simple one-dimensional differential equation~\eqref{eq:alphawithSpikes}, and observe that our model, despite its simplicity, finely accounts quantitatively for the number of pioneers. For instance, in the noise-induced oscillations regime, the model reflects the slow transitions arising in the system until the system reaches the critical proportion of pioneers $\alpha_c$, at which time a very sharp increase of $\alpha$ arises. 

In Fig.~\ref{fig:CorrectedAlpha}(B), we provide an extensive view of the dynamics of $\alpha$ as a function of $J$ and $\sigma$ in a four-dimensional representation (color represents the righthand side of equation~\eqref{eq:alphawithSpikes}). This representation recovers the eye-shaped noise-induced synchronization regime, corresponding to the parameter values associated with a uniformly positive value for intermediate values of noise and coupling.

\subsection{Conclusion: a subtle interplay of noise, connectivity and excitability}
This analysis of the trajectories provides a novel dynamical view of synchronization of stochastic particles. By finely dissecting the mechanisms of noise-induced synchronization in the FitzHugh-Nagumo network, we identified three distinct regimes of dynamics: synchrony for intermediate noise and coupling, flanked by asynchrony (large noise or low coupling) and clamping (low noise or high coupling). We provided a statistical physics account for these oscillations, and identified two key phenomena:
\begin{itemize}
	\item An asymmetry in the transition from rest to pioneer and reciprocally, leading to a spontaneous increase in the steady proportion of neurons in the excited state, in turn triggering
	\item A chain reaction leading all neurons to the excited state provided that the system reaches a sufficient proportion of neurons in that state.
\end{itemize}
These elements are not specific to the FitzHugh-Nagumo model, but arise in a wide class of excitable systems with noise; we will discuss in section~\ref{sec:Universality} their universality, illustrate the same property in distinct excitable systems, and further discuss the type of transition arising around the synchronized regime.

\section{Oscillations-Induced desynchronization and Parkinson's Disease}\label{sec:Desynchronization}
Studying the response of networks of excitable cells in the regime of spontaneous noise-induced oscillations is particularly relevant in the context of Parkinson's disease. Indeed, in Parkinson's disease, spontaneous oscillations emerge in the basal ganglia and motor cortex, typically in the beta band (10-30 Hz), and these arise while neurons show an increased excitability~\cite{degos,degos2}, together with enhanced electrical transmission~\cite{phookan2015gap,grace}. Some thirty years ago, it was observed that high-frequency stimulation  ($\sim$ 130Hz) in the basal ganglia relay nucleus, the subthalamic nucleus (Deep Brain Stimulation, DBS) had a remarkable effect of alleviating Parkinson's disease symptoms~\cite{degos2,benabid1987combined,lozano,ashkan,aum}. Clinically, the impact of DBS strongly depends on stimulation amplitude and frequency~\cite{momin}: too-high ($>180$Hz) frequency stimulation has been reported to be therapeutically ineffective~\cite{benabid1987combined,carvalho}, while low frequency stimulations is still debated: it has been shown to worsen some parkinsonian symptoms, such as tremors and rigidity~\cite{su}, possibly by imposing another oscillatory rhythm onto the basal ganglia network, while it appears beneficial for gait control and cognitive functions~\cite{dibiase-fasano}. A comparable dependence on amplitude was reported in patients~\cite{balaz}, with  too low stimulations abolishing the improvement of symptoms, sometimes with sudden, threshold-like loss of efficiency~\cite{conovaloff}. This observation led to us investigate in more detail the impact of high-frequency stimulation on noise-induced oscillations, and the impact of the stimulation parameters.

As shown in section~\ref{sec:NoiseInducedModel}, the origin of those oscillations is fundamentally stochastic (associated with random transitions between pioneer and resting regimes) and collective (chain reaction), and as such they are distinct in nature from more classical periodic systems. Periodically forced oscillators have a long history in the study of nonlinear dynamical systems, and the complex phenomena associated have been well described: they include phase locking with the stimulus, resonances, phase skipping and chaos, often associated with the presence of intricate dependences in amplitude and frequency of the forcing, as the classical Arnold tongues (see e.g.~\cite{keener}). Periodic forcing of noise-induced oscillations will induce a distinct phenomenology that we analyze below. 

\subsection{Periodic forcing in the noise-induced oscillations regime}\label{sec:DesyncObservations}
To emulate the impact of DBS on basal ganglia beta oscillations, we used balanced biphasic waves, typical DBS stimulation profiles advocated by Lilly in the 1960s~\cite{lilly} and widely used clinically. Such stimulations correspond to square waves of positive followed by negative currents with a zero mean:
\begin{equation}\label{eq:Lilly}
	I(t)=A \; H\left(\frac t T\right)
\end{equation}
where $A$ is the amplitude of the signal, $T$ is the stimulation period, and $H$ is a periodic profile square wave of period $1$ (we will typically use the sign of $\cos(2\pi t)$). The network equations with DBS stimulation thus read:
\[
\begin{cases}
	dv^i_t = \Big(f(v^i)-w^i+\frac J n \sum_{j=1}^n (v^j -v^i) + I(t)\Big)\,dt+\sigma dW^i_t\\
	dw^i_t = \eps(b\,v^i-w^i).
\end{cases}
\]

\begin{figure*}
  \centering
    \includegraphics[width=.9\textwidth]{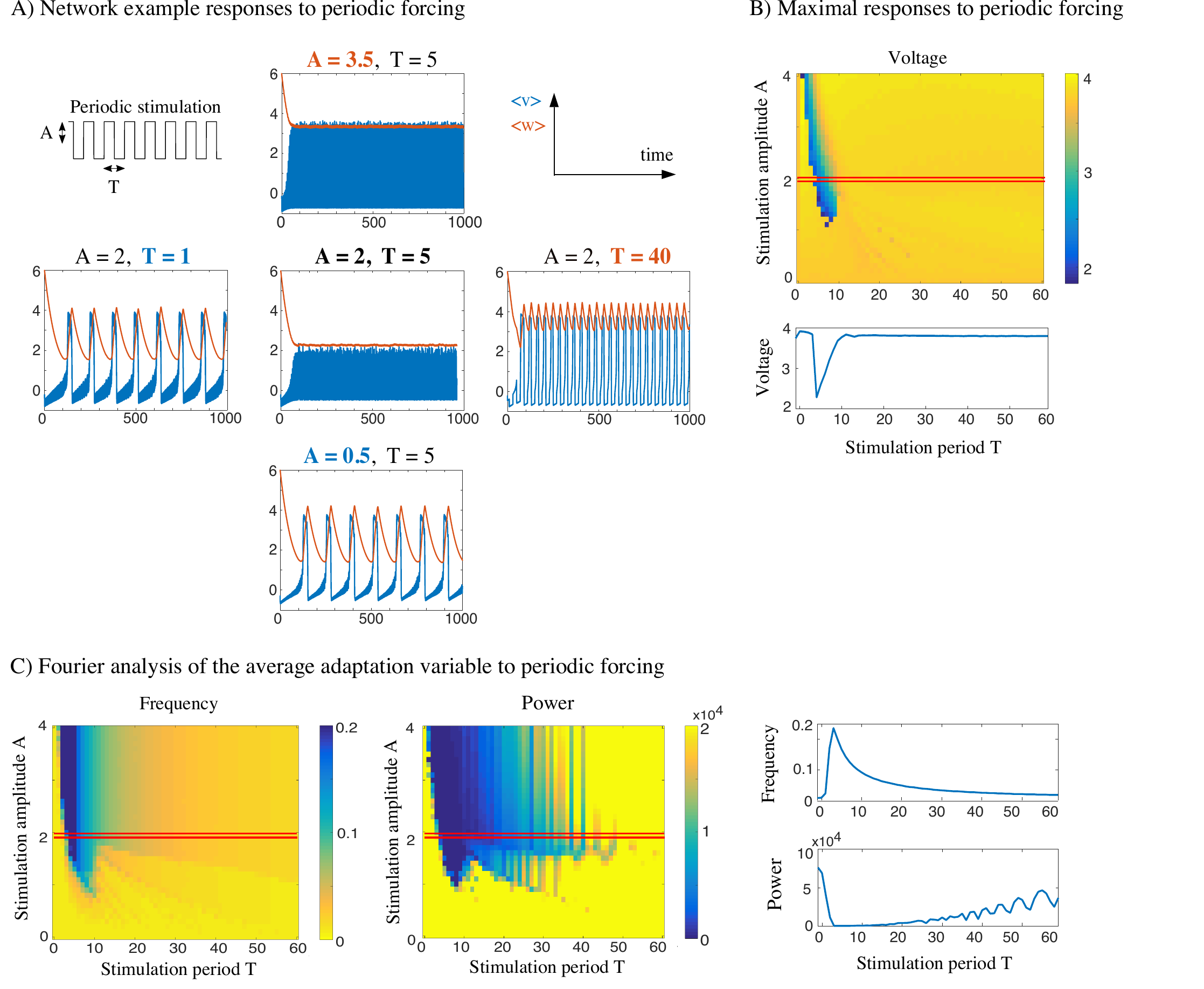}
  \caption{Desynchronization induced by high-frequency periodic stimulation of the Fitzhugh-Nagumo network equation (with same parameters used in Fig.~\ref{fig:phenomenon} and with $J=1.5$ and $\sigma = 1.5$) for various values of the amplitude $A$ and period $T$ of the stimulation current. (A) Trajectories and desynchronization: dynamics of the average voltage (blue) and adaptation variable (red). The system becomes desynchronized for relatively high-frequency stimulation (T=5), once the amplitude of the stimulation becomes strong enough to perturb the system. Increasing further the  stimulation frequency leads to the return of intrinsic oscillations (T=1), while slow periodic forcing induces a phase locking with the stimulation (T=40). (B) Top: The maximal value of the average voltage clearly delineates a region of desynchronization (in blue), corresponding to a dip in the maximal average as visible for fixed amplitude (bottom graph, corresponding to the red slice in the upper diagram). (C) Preferred frequency and maximal power of the Fourier transform of the average adaptation variable delimitates regions of synchronization (yellow), desynchronization (dark blue) and period skipping, with sections at fixed amplitude (red slices in the heatmaps and right panels). }
  \label{fig:DBS_inputFrequency}
\end{figure*}

Extreme regimes of stimulation frequency lead to two expected outcomes: very rapid periodic forcing has almost no impact on the spontaneous oscillations (Fig.~\ref{fig:DBS_inputFrequency}A, left), while very slow forcing locks network activity to the periodic signal (Fig.~\ref{fig:DBS_inputFrequency}A, right). 

Strikingly, for an intermediate value of periodic forcing frequency, the noise-induced oscillations are abolished (at a frequency relatively high compared to the spontaneous noise-induced oscillations frequency). In that anti-resonance regime, despite small amplitude oscillations of the voltage and adaptation variables in response to the DBS input, we observe a complete absence of collective dynamics or spiking (see Fig.~\ref{fig:DBS_inputFrequency}A, center). 

Stimulation amplitude has also an important impact: a too low amplitude $A$ barely affects the spontaneous activity (Fig.~\ref{fig:DBS_inputFrequency}, bottom). While it could be expected that increasing $A$ would lead the responses of the network to lock to the stimulation, we observed that, at the frequency tested, increasing amplitude did not alter the absence of oscillations for the parameters chosen. 

To quantify precisely the DBS amplitude and frequency associated with anti-resonance, we computed both the maximal value of the average voltage and adaptation (Fig.~\ref{fig:DBS_inputFrequency}B) and the maximal value of the Fourier transform of these variables (Fig.~\ref{fig:DBS_inputFrequency}C). We observed a significant drop both in the amplitude of the average voltage and adaptation, indicating the absence of collective spiking dynamics in the system, coinciding with a significant drop in the power-spectrum amplitude and frequency, highlighting the loss of synchrony at the network level. This loss of synchrony arises in a relatively wide range of parameter values and for a bounded band of frequency, within a region of parameters $(A,\omega)$ elongated along the amplitude axis. This shows that there exists an optimal stimulation frequency for desynchronization, as soon as the stimulation amplitude exceeds a threshold~\footnote{We note that for increasing stimulation frequency, the power-spectrum shows complex variations that combine the oscillations-induced desynchronization with more classical phase-locking phenomena. }.

\subsection{Statistical Mechanics of oscillations-induced desynchronization}\label{sec:DesyncModels}
We now discuss the statistical mechanics origin of this anti-resonance phenomenon. To this purpose, we extend the approach proposed for characterizing the statistical mechanics of noise-induced synchronization, and describe how the fraction of pioneers $\alpha$ is affected by periodic forcing. The application of a periodic input sweeps back and forth neurons from pioneer to rest and reciprocally, balancing dynamically two opposite phenomena:
\begin{itemize}
	\item during the \emph{excitation phase} (times for which $I(t)>0$), the rate of transition from rest to pioneer increases and the reciprocal rate decreases, neurons become more excitable (the resting state, if it persists under stimulation, gets closer from the separatrix) and $\alpha_c$ decreases. Consequently, neurons accumulate faster within the pioneer regime, and may trigger a spike faster than in the unperturbed system; 
	\item during the \emph{inhibition phase} (times for which $I(t)<0$), pioneers are rapidly brought back near the rest state. 
\end{itemize}
Because of these dynamical fluctuations of transition rates, the period of stimulation $T$ is crucial for anti-resonance. While a long inhibition phase would indeed prevent neuron from firing, the biphasic balanced stimulations profile will in turn present a long  (or large amplitude) excitation phase during which one or multiple spikes may be fired. 
In contrast, a too rapid signal will not allow enough time during the inhibition phase to drive back pioneers to rest, leading to a progressive accumulation of neurons in the pioneer state and eventually to a macroscopic spike, with largest frequencies having no impact on the period. At intermediate frequencies, the two phenomena may balance and jam the system below the chain reaction threshold. For this to occur, the period of the stimulation should be smaller than the time it takes for the system to reach the chain reaction threshold when the input is $+A$, but not much smaller so as to allow a proper compensation during the inhibition phase. A stimulation frequency higher than spontaneous oscillations frequency is needed to prevent the emergence of synchrony, in line with DBS in Parkinson's disease. 

\begin{figure}
  \centering
    \includegraphics[width=\textwidth]{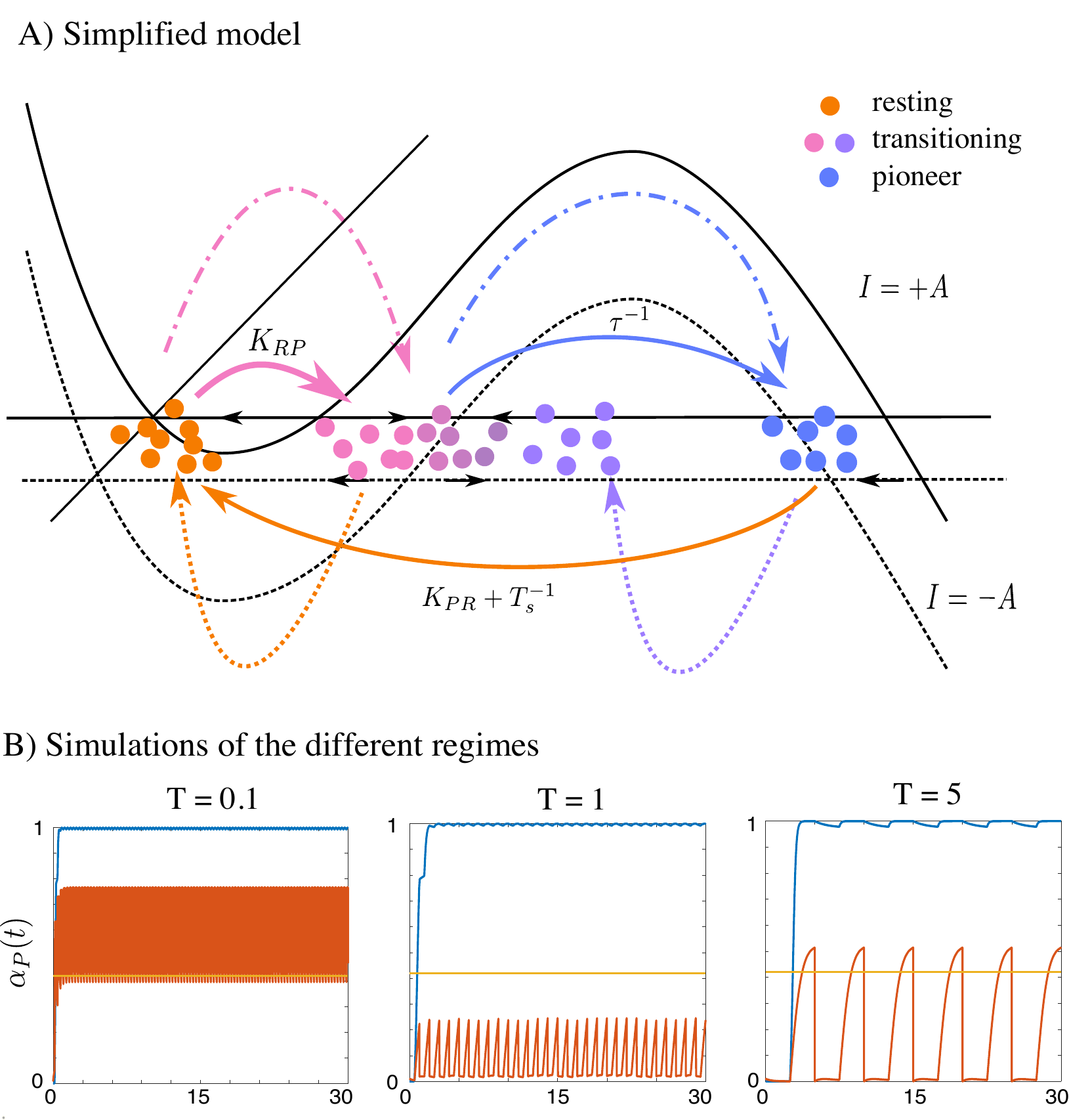}
  \caption{Principle of the model of oscillations-induced desynchronization. (A) Phase-plane of the toy model, with nullclines for $I=A$ (solid) or $I=-A$ (dotted), together with the three populations of cells: resting (orange), transitioning (pink-to-violet according to the value of $v$) and pioneer (blue), together with the transitions at fixed input (solid lines) and at switching times from $+A$ to $-A$ (dashed lines) or from $-A$ to $+A$ (dot-dashed lines). (B) Simulation of the proportion of pioneers in the rest-pioneer model (blue) and in the three-population model (red) as a function of time for increasing stimulation period. No qualitative change is observed in the rest-pioneer model (blue lines): for any frequency, a chain reaction is predicted. In the three-population rest-transitioning-pioneer model, we recover the phenomenon of Fig~\ref{fig:DBS_inputFrequency}. (left): for rapid oscillations, compensation of the pioneers does not occur, and the system eventually exceeds $\alpha_c$ (yellow line) (middle): optimal oscillation frequency: the system stabilizes around the critical value with a mean below that value preventing spiking, and (right): slow oscillations synchronize the system to the input. In that model, $K_{RP}(\pm A,\cdot)$ and $K_{PR}(\pm A,\cdot)$ were computed numerically as in the previous section, with $A=2$, $\tau=2$ and the probability to switch is a sigmoidal ($\tanh$) function of $(\tau\omega)^{-1}$ with slope $4$ and threshold $0.2$, and the periods of stimulation chosen are, from left to right, $T=0.1,\; 1,\; 5$. }
  \label{fig:Model_DBS}
\end{figure}

Characterizing quantitatively the anti-resonance phenomenon requires keeping track of neurons continuously switching from rest to pioneer. We thus extended the simple resting-pioneer model to include \emph{transitioning} neurons: instead of instantaneous transitions from rest to pioneer, neurons that cross the separatrix enter a transitioning state, before gradually turning into pioneers in a time evaluated as the typical time of transition (see Fig.~\ref{fig:Model_DBS}-A). Stochastic transitions from rest to transitioning neuron and from pioneer to rest occur as characterized in the previous section, and the rates now depend on time following the fluctuations of the input $I(t)$. In addition to these stochastic transitions, deterministic transitions from pioneer to rest through spiking, and from transitioning neurons to pioneers according to the flow, occur at typical times respectively denoted $T_s$ and $\tau$. 

In the periodically forced system, the resting, transitioning and pioneer states are relative to the value of the input. Moreover, in the stochastic network system, the population of neurons do not have homogeneous voltages (in particular because of noise and continuity of the trajectories). This variation in voltage can no more be neglected in the anti-resonance phenomenon. For instance, when the input switches from $+A$ to $-A$, a fraction of transitioning neurons (including in particular those that have just switched from rest) will have low voltages within the resting range of the system with $I=-A$, and a fraction of neurons from the pioneer population will belong to the transitioning range of voltage. The longer a transitioning neuron has been in the transitioning state (with, say, $I=+A$), the larger its voltage, and thus the less likely it will become a resting neuron when the input switches to $-A$. Similarly, when the input switches from $-A$ to $+A$, some resting neurons become transitioning and some transitioning neurons become pioneers. 

We developed a simple toy model recapitulating these phenomena in a two-dimensional equation, describing the fractions of: (i) \emph{resting} neurons, that did not start their transition, (ii) \emph{transitioning} neurons, that initiated a transition to pioneers, and (iii) \emph{pioneers}. Respective proportions in each state are denoted $\alpha_R (t)$, $\alpha_{T}(t)$, $\alpha_{P}(t)$ (with, for all times, $\alpha_R (t)+\alpha_{T}(t)+\alpha_{P}(t)=1$), and evolve according to the following dynamics (see Fig.~\ref{fig:Model_DBS}-A):

\begin{itemize}
	\item A resting neuron becomes a transitioning neuron with rate $K_{RP}(I(t),\alpha_P(t))$ (transition rate with making explicit the dependence in $I$, solid pink arrow in Fig.~\ref{fig:Model_DBS}-A);
	\item A transitioning neuron becomes pioneer at rate $\frac 1 \tau$ (solid blue arrow in Fig.~\ref{fig:Model_DBS}A);
	\item A pioneer neuron returns to rest due to noise with rate $K_{PR}(I(t),\alpha_P(t))$, and after spiking with a rate $1/T_s$ (orange arrow in Fig.~\ref{fig:Model_DBS}A);
	\item When the input switches from $I=+A$ to $-A$, a proportion $S_+(T/\tau)$ of transitioning (resp. pioneer) neurons become resting (resp. transitioning) neurons, with $S_+$ a decreasing sigmoid with $S_+(0)=1$ and $S_+\to 0$ at infinity (dotted orange and purple arrows in Fig.~\ref{fig:Model_DBS}A);
	\item When the input switches from $I=-A$ to $+A$, a proportion $1-S_-(T/\tau)$ of transitioning (resp. resting) neurons become pioneer (resp. transitioning) neurons, with $S_-$ a decreasing sigmoid with $S_-(0)=1$ and $S_-\to 0$ at infinity 
	(dot-dashed pink and blue arrows in Fig.~\ref{fig:Model_DBS}A)).
\end{itemize}
Therefore, during periods of constant input, the fractions of pioneers, resting and transitioning neurons satisfy the equations:
\[
\begin{cases}
	\dot{\alpha}_P=-\alpha_P \; K_{PR}(I(t),\alpha_P(t)) + \frac{\alpha_T}{\tau} -\frac{\alpha_P}{T_s}\\
	\dot{\alpha}_T=\alpha_R  \; K_{RP}(I(t),\alpha_P(t)) - \frac{\alpha_T}{\tau}\\
	\alpha_R  =1-\alpha_P-\alpha_T,
\end{cases}
\]
together with the jumps when the input switches:
\[
 (I:\,-A \to +A)
 \begin{cases}
	\alpha_P=\alpha_P+S_+(\frac{T}{\tau}) \alpha_T\\
	\alpha_T=S_+(\frac{T}{\tau}) \alpha_R +(1-S_+(\frac{T}{\tau})) \alpha_T
\end{cases}
\]
\[
 (I:\,+A \to -A)
 \begin{cases}
	\alpha_P=S_-(\frac{T}{\tau}) \alpha_P\\
	\alpha_T=S_-(\frac{T}{\tau}) \alpha_T +(1-S_-(\frac{T}{\tau})) \alpha_P
\end{cases}
\]

We numerically simulated this simple system of ODEs with jumps using the rates of transitions and spike duration computed in section~\ref{sec:NoiseInducedModel}, and found that it accurately recapitulates all observations associated with the oscillations-induced desynchronization (Fig.~\ref{fig:Model_DBS}-B). In particular, we recover the non-monotonic dependence in stimulation frequency. At high frequency, the input plays a minor role: resting neurons progressively transition to pioneer, and neurons end up firing collectively, since the too brief inhibition phase is inefficient to counterbalance the accelerated transition of resting neurons to the pioneer state (Fig.~\ref{fig:Model_DBS}-B, left). At low frequency, macroscopic spikes are fired, as visible in the pioneer fraction exceeding the critical fraction (Fig.~\ref{fig:Model_DBS}-B, right). However, intermediate frequencies can finely balance the transitions to the pioneer state and stabilizing the fraction of pioneers below the critical fraction $\alpha_c$. In that case, a majority of neurons end up fluctuating within the transitioning regime, and the fraction of pioneers does not reach the critical fraction associated with the chain reaction (Fig.~\ref{fig:Model_DBS}-B, middle).

This model allows a deeper understanding of some features of the oscillations-induced desynchronization. Starting from a regime of noise-induced synchronization, it is clear that to prevent any firing, a necessary condition is that the input is large enough for the system with input $-A$ to be clamped to rest. Indeed, our 3-variables model alternates tracking the high and low equilibria of $\alpha_P$, denoted $\alpha^+$ and $\alpha^-$ as $I$ switches from $A$ to $-A$. At low stimulation amplitude, $\alpha^+$ and $\alpha^-$ are close from the equilibrium value of $\alpha$ in the absence of input, and therefore both fractions will be above $\alpha_c$ for sufficiently low amplitude. In that case, no desynchronization will occur, as was observed in the full stochastic system (Fig.~\ref{fig:DBS_inputFrequency}(B)). Moreover, a decreasing relationship between $A$ and $\omega$ was observed in the regime of desynchronization. From the simplified model viewpoint, this relationship can be interpreted noting that, for larger input amplitudes, the proportion of pioneers will more quickly increase above $\alpha_c$, requiring faster switches to stabilize below $\alpha_c$. 

\subsection{Information Capacity}\label{sec:Information}
Stochastic excitable systems thus reproduce the alteration of spontaneous oscillations in the presence of high frequency stimulation as observed in Parkinson's disease. Taking the parallel with Parkinson's disease one step further and the oscillation-induced desynchronization regime analogy with the therapeutic effect of DBS, we investigated the possible impact of periodic stimulation on the information capacity of the network. Indeed, DBS has been associated with the restoration of normal activity patterns in the basal ganglia and a decrease in motor symptoms in parkinsonian patients, as discussed above.  

In the stochastic system, when oscillations abolish collective noise-induced desynchronization, the system is maintained in a dynamical state balancing the natural tendency of the system to fire in the noise-induced oscillations regime and keeping the system on the verge of a chain reaction. Remaining within a highly reactive state, the stochastic network may thus be able to respond rapidly and precisely to external stimulations. 

\begin{figure*}[h]
  \centering
    \includegraphics[width=.8\textwidth]{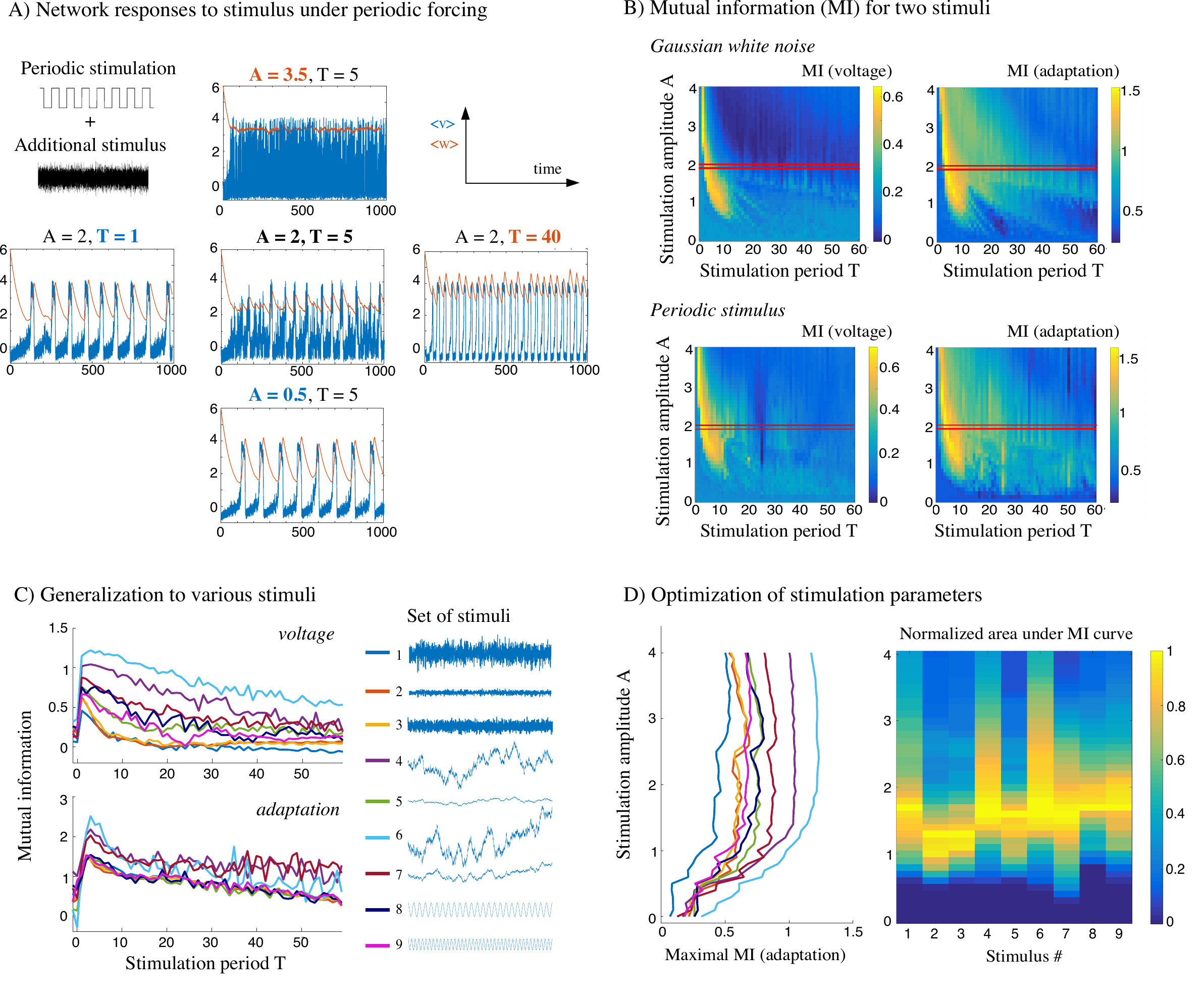}
  \caption{Information capacity of the Fitzhugh-Nagumo network model, parameters as in Fig.~\ref{fig:Model_DBS}A (oscillations-induced desynchronization, with $n=1\,000$ neurons). An external stimulus was added to the system, taking one of 9 forms as depicted in the bottom-left panel and labeled 1-9 and color-coded (1-3: Gaussian white noise, 4-7: Ornstein-Uhlenbeck processes, 8-9: sine waves). (A) Trajectories of mean network activity in response to Gaussian white noise (stimulus 3), as a function of DBS parameters (amplitude and frequency). (B) Mutual information $\eta$ for stimulus 3 (top) and 9 (bottom) calculated for different values of the amplitude $A$ and period $T$ of the periodic stimulation from the voltage or adaptation variables. (C) Summary of the mutual information for all nine stimuli as a function of the period $T$ of the periodic stimulation ($A = 2$). (D) Maximal value of the mutual information found for each stimulation amplitude (left) and area under the mutual information curve (above a threshold equal to twice the basal mutual information level without periodic forcing), normalized by the largest area within each stimulus. These indexes were calculated from the voltage responses.
}
  \label{fig:entropy}
\end{figure*}

To test this hypothesis, we computed the mutual information of the periodically forced FitzHugh-Nagumo stochastic network responses to the presence of additional stimuli. These stimuli were injected as a current, thus affecting the voltage variable $v$, of each neuron of the network. To evaluate the information capacity of the network, we computed the entropy of the responses in the voltage or recovery variables. To this purpose, we considered the response of each neuron as a sample realization of the process under consideration. Rigorously, while those neurons are not independent because of the interaction term, the \emph{propagation of chaos} (Boltzmann's molecular chaos) property ensures that, in the large $n$ limit, neurons become statistically independent realizations of the same process. To avoid any bias arising due to the regularity of the spontaneous response patterns (too high frequency) or to the forcing (too low frequency), we calculated how input modified those patterns by computing entropies in a time window of size $\Delta t$ evaluated according to the period of the network responses in the absence of stimulus (period evaluated through the Fourier transform of the solutions, see Fig.~\ref{fig:entropy}).

{We thus computed a \emph{stimulus-specific windowed entropy} $H_s$ as an averaged entropy across time windows $[k\Delta t,(k+1)\Delta t)$ by computing the probability distribution of the voltage or adaptation variables over all neurons within a given window in response to a given stimulus. In detail, we numerically computed the solution of the stochastic FitzHugh-Nagumo equation $(v^i(t),w^i(t))$ for $t\in [0,K\,\Delta t]$ with $K\in \mathbb{N}\setminus \{0\}$ and $i\in\{1,\cdots,n\}$. Given a voltage or recovery resolution $\Delta B_{v,w}$ (chosen to be $0.025$ in our simulation), we define fixed partitions of the voltage and recovery variables $\{v_1,\cdots,v_{M_1}\}$ (voltage partition) and $\{w_1,\cdots,w_{M_2}\}$, and computed a discretized empirical probability distributions $\hat{p}_{v,w}^\alpha(k)$ as the probability to find a neuron in the network with a voltage (or recovery variable) within the segment $[v_{\alpha},v_{\alpha+1}]$ 
(respectively, $[w_{\alpha},w_{\alpha+1}]$) during the time interval $[k\Delta t,(k+1)\Delta t)$ with $\alpha \in\{1,\cdots, M_1-1\}$ and $k\in \{0,\cdots, K-1\}$. The stimulus specific entropy was computed as the associated Shannon's differential entropy:
\[H_s^{v}=\frac{1}{Z} \sum_{k=0}^{K-1} \sum_{\alpha=1}^{M_1-1} \hat{p}_{v}^\alpha(k)\log\left(\frac{\hat{p}_{v}^\alpha(k)}{\Delta B_{v}}\right) \]
(a similar expression is used for the adaptation entropy), where $Z$ is a normalization constant (generally $K M_{1,2} \Delta B_{v,w}$) that acts as a simple scaling term with no impact on the qualitative results.}

{Similarly, the \emph{averaged windowed entropy} $H_{all}$ was calculated as the average of the stimulus-specific entropies across all stimuli considered, and the mutual information was then defined for each stimulus as:
\[\eta = H_{all} - H_s.\] }

The stimulus set consisted of nine stimuli: three centered Gaussian white noise with distinct standard deviations, four centered Ornstein-Uhlenbeck processes and two sine waves with identical amplitude and distinct frequencies (top right panel, Fig.~\ref{fig:entropy}). {To accurately compare entropies and obtain results that do not depend on the particular realization of noise, we froze the intrinsic noise realizations and used the same trajectories of white noise for all conditions (we used 5 independent simulations for each condition; results were all consistent; the graphs show the averaged entropies across the noise realizations).} Entropies were computed based on the empirical distributions on a fixed grid (step: $0.025$, voltage from $-2$ to $6$, adaptation from $0.5$ to $5.5$). 

As can be seen from the average voltage traces (Fig.~\ref{fig:entropy}A), in the desynchronized regime, the network shows an increased `reactivity' to the stimulus (in this case, Gaussian white noise) and a reduced dependence on the periodic DBS forcing pattern. This situation contrasts with the regimes in which intrinsic or forced high-amplitude oscillations dominate even in the presence of the stimulus. Quantitatively, this results in a significant increase in the mutual information, spanning the entire band characteristic of the desynchronized regime (Fig.~\ref{fig:entropy}B), when quantified from both the voltage and recovery variables. This significant elevation of the mutual information in the desynchronized regime can be observed for every stimulus considered, regardless of their nature or their intensity (Fig.~\ref{fig:entropy}C). In addition, the mutual information and the amplitude of oscillations are correlated to the stimulation frequency: indeed, the sharp decay of oscillations as the frequency of the periodic forcing is decreased parallels with a sharp rise of the mutual information, while as the regime moves into periodically-forced oscillations, the mutual information decreases smoothly. The mutual information between the stimuli and the network responses under low-frequency stimulation also tends to be close to the mutual information characteristic of the intrinsically oscillating regime.  

The mutual information varies with the amplitude of the periodic stimulation: indeed, as the amplitude of the periodic stimulation increases, there is an increase in the peak of the mutual information (Fig.~\ref{fig:entropy}D), before saturating. Yet, the width of this peak in mutual information has a non-monotonic relation with the stimulation amplitude: low-amplitude stimulation yields a medium improvement in stimulus encoding over a wide range of frequency, while for high-amplitude stimulation, the increase in mutual information occurs on an increasingly narrow frequency band (as can be seen for two stimuli in Fig.~\ref{fig:entropy}B). Therefore, for each stimulus, an optimal stimulation amplitude can be defined, such that the area under the mutual information curve in the desynchronized regime is maximized (Fig.~\ref{fig:entropy}D): this optimal amplitude is relatively low, and consistent for all stimuli tested, indicating the presence of an optimal forcing for information processing regardless of the stimulus type.

Overall, these results suggest that highly stable intrinsic or forced oscillations limits stimulus information encoding, while their destabilization through high-frequency stimulation endows the system with more computational capabilities, and that the gain in information capacity is maximized for appropriate stimulation amplitude and frequency. 

\section{Universality }\label{sec:Universality}
We now study the universality of the noise-induced synchronization and anti-resonance in a class of interacting excitable elements, as well as the existence of universal transitions by which synchrony emerges from clamping or asynchrony. 
We give further evidence of this universality by presenting numerical simulation of three other stochastic neural network models and bifurcations analyses.

\subsection{How do noise-induced oscillations emerge and disappear?}

\subsubsection{Noise-induced oscillations}
Section~\ref{sec:NoiseInducedModel} has shown, in the FitzHugh-Nagumo model, that noise-induced oscillations were the result of the buildup of an imbalance in the stochastic transition rates of neurons between the resting state and an excited state eventually leading to a chain reaction, and that this phenomenon could be described by a simple one-dimensional birth-and-death Markov chain (or its ODE counterpart) tracking the fraction of cells in each state. This phenomenon relies on the following few crucial elements:
\begin{itemize}
	\item Excitability, accounting for the transitions between rest and an excited state;
	\item Confining-like interactions (e.g., diffusive coupling), preserving the coherence of the elements.
\end{itemize}
For excitable systems with such interactions, we conjecture that one will recover both elements leading to noise-induced oscillations. In particular, for such systems, the existence of a critical fraction of excited neurons above which a synchronized spike occurs will be ensured provided that the coupling strength is sufficiently large to prevent individual spikes. Indeed, given a particle near the resting state, the presence of a large fraction of excited neurons and confining interactions will act as a driving force counterbalancing the stability of the resting state, and, for coupling or $\alpha$ large enough, will lead to a destabilization of the resting state. Moreover, in such systems, a proper amount of noise will naturally lead the system to exceed that critical fraction of excited neurons. Indeed, the transition rates from rest to the excited state and reciprocally are generally asymmetric in excitable systems. This asymmetry could be associated to the fact the excited state is not a stable equilibrium of the dynamical system, but a transient state preceding a large excursion away from the resting state, while the dynamics near the resting state are stationary. In other words, excited elements are driven to a remote state and less likely to return to rest than elements near rest that have a vanishing vector field. This should naturally lead to an imbalance in the rates of transition: once a neuron has transitioned to the excited state, it will likely have a smaller rate of transition to rest than the reciprocal rate. Moreover, confining interactions will naturally amplify this imbalance. For this collective phenomenon to build up, a sufficiently strong coupling is necessary to provide coherence to the set of neurons, but a too large coupling will limit the transitions, clamping the system at rest. Therefore, we conclude that stochastic networks of excitable systems should present noise-induced oscillations for a limited range of coupling strengths allowing sufficient coherence to the set of elements, yet sufficient flexibility to allow the buildup of the chain reaction. Noise levels should also be limited, as  observed in the FitzHugh-Nagumo network:
	\begin{itemize}
		\item Too little noise leads to rare transitions, and because of the natural return of each element to rest, the critical proportion of excited elements is never reached;
		\item Too much noise, making the transition rates from rest to pioneer and reciprocally more symmetric and breaking the excitable structure of the intrinsic dynamics, will prevent any significant increase in the fraction of excited elements.
	\end{itemize}

These phenomena are thus not specific of the FitzHugh-Nagumo network, and should be valid in a broad class of excitable systems with synchronizing coupling. 

\subsubsection{Transitions to and from noise-induced oscillations}
As observed in the FitzHugh-Nagumo network for a given coupling level, three regimes arise as noise is increased: clamping near the resting state, oscillations and asynchrony. The nature of the transitions to and from the oscillatory regime can also be inferred from a statistical mechanics analysis, and shall thus share the same universality properties. 

At low noise, the rarity of transitions between rest and excited states prevent the system from reaching the critical proportion of excited elements, and thus, in the long run, the system reaches a steady proportion of excited elements below that critical threshold. As noise is increased, transition rates from rest to pioneer and reciprocally will increase asymmetrically, leading to a progressive increase in the steady proportion of pioneers, until that proportion reaches the critical fraction associated with the chain reaction. In the vicinity of this transition and within the clamping regime, the proportion of pioneers asymptotically tangents a level slightly below the critical fraction, and as the transition is crossed, it takes an arbitrary long time to trigger a synchronization event; moreover, this event will be massively synchronized because of the relatively low level of noise. Therefore, as noise is slowly increased from the clamped regime, the collective dynamics suddenly transition from a stationary to arbitrary slow, large amplitude highly synchronized oscillations. This type of dynamics is evocative of homoclinic bifurcations, and we conjecture that excitable networks displaying noise-induced transitions switch to these oscillations through a collective homoclinic-like bifurcation. 

When noise is further increased, oscillations will reach a finite frequency associated with the deterministic time needed for the population of excitable elements to reach the chain reaction threshold, while losing progressively coherence due to an increased influence of independent fluctuations. This loss of coherence will lead to a decrease in the average voltage oscillation amplitude, while the period of the oscillations remains lower-bounded by the typical time of an excursion. Therefore, as noise is increased, oscillations will progressively disappear through a desynchronization evocative of a supercritical Hopf bifurcation. 

This explanation, perfectly in line with the numerical evaluation of the mean and period of the oscillation in the FitzHugh-Nagumo network in Fig.~\ref{fig:phenomenon}, will be confirmed in several excitable systems in section~\ref{sec:UniversalityExamples}. Moreover, we will verify the presence of the conjectured transitions in the two simple models (theta neuron and Wilson-Cowan models) for which one can access the bifurcation diagram of the probability distribution. 

\subsubsection{Oscillations-induced desynchronization}
In the class of excitable systems showing noise-induced oscillations, we further tested the impact of high-frequency periodic forcing, and found again that the phenomenon enjoys a relatively broad universality. The statistical mechanics of the phenomenon uncovered in the FitzHugh-Nagumo network in the previous section indeed calls upon relatively general mechanisms resulting in neurons remaining dynamically within a transitioning regime whereby neurons are neither at rest, nor fully in the excited regime: positive phases of the input, while they may accelerate the emergence of a chain reaction, are quickly compensated by the negative phases of the input. This general phenomenon requires however a relatively slow transition from rest to pioneer, which is not always found in simple models. We will recover this desynchronization in other models in section~\ref{sec:UniversalityExamples}.

\subsection{Exploring the universality class}\label{sec:UniversalityExamples}
To confirm the universality of both transitions, we simulated three other networks of excitable elements with noise, in the large $n$ limit, as coupling and noise are varied. 

\subsubsection{Morris-Lecar model} 
We start considering the stochastic, electrically coupled network of Morris-Lecar neurons~\cite{morris-lecar}, a classical biophysically realistic neuron model particularly interesting for its relative simplicity yet direct relationship with electrophysiology and with the Hodgkin-Huxley model. In this model, the state of neuron $i\in\{1\cdots n\}$ is described by a voltage variable $v^i$ and an adaptation $w^i$ whose dynamics are governed by the equations:
\[\begin{cases}
dv^i_t&=\frac 1 c \Big(I-g_{Ca}\,(v^i-v_{Ca})\,m_{\infty}(v^i)-g_K\,(v^i-v_K)\,w\\
&-g_L\,(v-v_L) + I + J (\langle v\rangle -v^i)\Big)\,dt+\sigma dW^i_t\\
dw^i_t&=\frac{\phi}{\tau_w(v^i_t)} (w_{\infty}(v^i_t)-w^i_t).
\end{cases}\]
with $\langle v\rangle(t)=\frac 1 n \sum_{j=1}^n v^j_t$. In that model, $c$ denotes the membrane capacitance, $I$ is a current, $g_L,\,g_{K}, g_{Ca}$ are the leak, $K^+$ and $Ca^{2+}$ conductances through membrane channels, $v_{l}, v_{K}, v_{Ca}$ their reversal potentials, $m_{\infty}(v)$ accounts for an instantaneous calcium current, $\phi$ is a reference frequency, $\tau_w$ the timescale of adaptation $w$ and $w_\infty$ is the quasi-steady-state value of $w$. We refer to~\cite[Chap. 3.2]{ermentrout-terman:10b} for the specific sigmoidal shapes of $\tau_w$, $m_\infty$ and $w_\infty$ as well as for basic parameter values. The above equation incorporates noisy currents driven by independent Brownian motions $(W^i_t)$ and diffusive coupling modeling electrical synapses. 

\begin{figure*}
  \centering
    \includegraphics[width=1\textwidth,height=20cm,keepaspectratio]{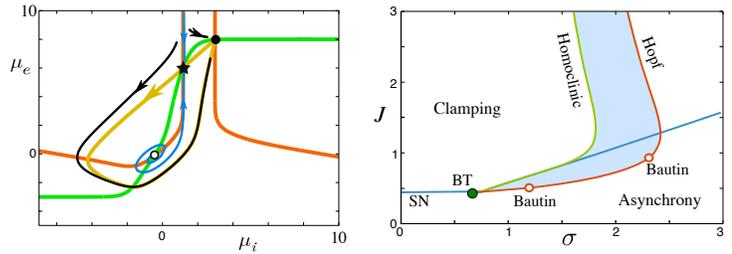}
  \caption{Universality of the transitions: (A) \textbf{Morris-Lecar model}: Noise-induced transition and oscillations-induced desynchronization; parameters as in~\cite[Table 3.1, SNLC]{ermentrout-terman:10b}, with $\phi=0.02$, $V_1=-1.3$, $V_4=10$ and basic applied current $I=35$. (A1) shows the noise-induced synchronization phenomenon as a function of coupling strength $J$ and noise $\sigma$; trajectories: 20 sample voltage traces (dark blue) and mean (bold cyan); heatmaps: Statistics of the synchronization via Fourier transform peak period (top) and amplitude (bottom) as a function of coupling strength and noise. (A2) Responses of the system within the noise-induced oscillations regime ($J=1.5$, $\sigma=3$) to a Lilly pulse eq.~\ref{eq:Lilly} for various frequencies and amplitudes; at appropriate frequency and amplitude, oscillations disappear (center); trajectories: 20 sample adaptation traces (red) and mean (bold black); heatmap: amplitude of the Fourier transform as a function of period and amplitude of the stimulation. (B) \textbf{The theta neuron} with $a=0.04$. Two-parameter bifurcation diagram of the Fokker-Planck equation of the theta neurons as a function of coupling strength and noise shows a saddle-node bifurcation manifold (blue) and a Hopf bifurcation manifold (red), together with two codimension-two cusp and two Bogdanov-Takens bifurcations, associated with a branch of saddle-homoclinic bifurcation (yellow). The Hopf and homoclinic bifurcation curves delineate an eye-shaped noise-induced synchronization regime (sky blue region). (C) \textbf{The excitatory/inhibitory Wilson-Cowan mean-field system} with $g_{ee}=15$, $g_{ei}=-12$, $g_{ie}=16$, $g_{ee}=-5$, $I_e=0$, $I_i=-3$, $S(x)=\erf(3\,x)$. The system exhibits an intrinsic excitable structure (left) with a single stable fixed point (black circle) and a saddle fixed point (black star) whose stable (blue) and unstable (yellow) manifolds organize the excitability. Green and red curves are the nullclines of the system. (Right) Two-parameter bifurcation diagram of the excitatory/inhibitory Wilson-Cowan system features a Hopf (red; solid: supercritical, dashed: subcritical, red circle: codimension-two Bautin bifurcation) and a Saddle-Node (blue) bifurcations colliding at a Bogdanov-Takens bifurcation (green BT circle), from which point emerges a homoclinic bifurcation (green line). Near the BT point, the saddle-node bifurcation shows a cusp (not visible in the diagram). The homoclinic and Hopf bifurcations delineate a region of noise-induced oscillations (blue), that extends slightly beyond the Hopf curve between the two Bautin bifurcations and disappear through a fold of limit cycles (not shown). As noise is increased, oscillations emerge from clamped regimes through the homoclinic bifurcation and disappear through the Hopf bifurcation (or the fold of limit cycles) into an asynchronous state. }
  \label{fig:Universality}
\end{figure*}

It is well known that this system is excitable within a wide range of parameter values~\cite{ermentrout-terman:10b}. We thus analyzed the impact of noise and electrical coupling within this excitability regime, and recovered the noise-induced oscillation transition, similar in many ways to the observations made in the FitzHugh-Nagumo network (see Fig.~\ref{fig:Universality}A). In particular, at high coupling or low noise, the system is clamped in the vicinity of the resting state (left and top diagrams in Fig.~\ref{fig:Universality}A1), while at low coupling or large noise, asynchronous dynamics take over (bottom and right diagrams in Fig.~\ref{fig:Universality}A1). Between these two regimes, noise induces perfectly periodic synchronized oscillations. The type of transitions is also clearly recovered: for a fixed level of coupling, sharp and large amplitude, arbitrarily slow oscillations appear suddenly as noise is progressively increased, again evocative of a homoclinic transition, and, as noise is further increased, these oscillations progressively lose synchrony, as visible in the gradual decrease in the amplitude of the average voltage at reaching the asynchronous regime. These observations are also visible in the Fourier analysis associated. 

Furthermore, the noise-induced oscillations were found to disappear under application of a biphasic Lilly pulse $I(t)=A H(t/T)$ (eq.~\eqref{eq:Lilly}) for periods and amplitudes within specific bounds: too rapid oscillations do not affect the spontaneous oscillatory dynamics 
(Fig.~\ref{fig:Universality}A2, left),  too slow oscillations lock the system to the stimulus (Fig.~\ref{fig:Universality}A2, right), and appropriate frequency and amplitude abolish the synchronization (Fig.~\ref{fig:Universality}A2, center), a phenomenon arising in a relatively broad range of stimulation periods and amplitudes as shown in the Fourier analysis presented in the lower-left panel of Fig.~\ref{fig:Universality}A2.

\subsubsection{The electrically coupled theta neuron network} 

The theta neuron constitutes a canonical example of excitable system~\cite{ermentrout-terman:10b}. The stochastic electrically coupled theta neuron system describes the phase of neuron $i$ in a $n$-neurons network as a variable $\theta^i\in \mathbb{S}^{2\pi}$ (the 1-dimensional torus $\R/2\pi\mathbb{Z}$) through the equations:
\begin{multline*}
	d\theta^i_t=\Big[1-\cos(\theta^i_t)+(1+\cos(\theta^i_t))\Big(-a - \sigma^2 \sin(\theta^i_t)\\
	+\frac{J}{n}\sum_{j=1}^n \left(q(\theta^j_t)-q(\theta^i_t)\right)\Big)\Big]\,dt
	+ \sigma\,(1+\cos(\theta^i_t))\,dW^i_t
\end{multline*}
with $q(\theta)=\frac{\sin(\theta)}{1+\cos(\theta)+\eps}$ for $\eps$ small (in our numerical simulations, $\eps=0.001$). Classical theory of mean-field limits ensures that, for $n\to \infty$, the probability distribution $p(t,\theta)$ of any given neuron in the network to be at phase $\theta$ at time $t$ converges to the solution of the non-local equation:
\begin{align}
\nonumber	\partial_t p =& -\partial_\theta\Big[p\,\Big(1-\cos(\theta)+(1+\cos(\theta))\,\Big(-a\\
\nonumber	&- \sigma^2 \sin(\theta)+J\,\big(\int_{0}^{2\pi} q(\theta')p(t,\theta')\,d\theta'-q(\theta)\big)\Big)\Big)\Big]\\
\label{eq:MFtheta}	&+\frac{\sigma^2}{2}\partial_\theta^2 (p(t,\theta)(1+\cos(\theta))^2).
\end{align}
Numerical simulations of the stochastic network equation as well as the mean-field Fokker-Planck equation recover the emergence of synchronized oscillations for appropriate coupling and noise levels. Here, instead of presenting numerical simulations of the network equation, we computed numerically the bifurcations of the mean-field equation~\eqref{eq:MFtheta}, thus finding the precise boundaries of the synchronization regime.

To this end, we discretized this equation for $\theta\in\{\theta_k=\frac{k 2\pi}{N_{grid}}, k=0\cdots N_{grid}-1\}$ with $N_{grid}=100$, thus replacing the non-local partial differential equation into a $N_{grid}$-dimensional ordinary differential equation describing the probabilities $p_k(t)=p(t,\theta_k)$, similar to the system written above, with the operator $\partial_\theta (f)$ ($f$ is here a dummy variable) discretized as a centered finite-difference $(f(\theta_{k+1})-f(\theta_{k-1}))/2\delta$, the operator $\partial_\theta^2 (f)$ by $(f(\theta_{k+1})+f(\theta_{k-1})-2f(\theta_k))/\delta^2$, and the integral term $\int_0^{2\pi} f(\theta')d\theta'$ by $\sum_{k=0}^{N_{grid}-1} f(\theta_k)\delta$, with $\delta=2\pi/N_{grid}$. 

The two-parameter bifurcation diagram of this equation as a function of the coupling strength and noise are depicted in Fig.~\ref{fig:Universality}(B). The bifurcation diagram is organized around two codimension-two Bogdanov-Takens bifurcations, sharing the same Hopf, homoclinic and saddle-node bifurcation manifolds (two cusp bifurcations were also found and have no impact on the noise-induced oscillations phenomenon).  We recover, as in the FitzHugh-Nagumo or the Morris-Lecar system, an eye-shaped region of noise-induced oscillations for intermediate values of noise and coupling, splitting the parameter space between clamping and asynchronous regimes. 

We emphasize that the possibility to access the bifurcation diagram for the probability distribution of the Mean-Field Fokker-Planck equation allows supporting the conjecture related to the type of bifurcations surrounding the noise-induced oscillations regime. We indeed find that the homoclinic bifurcation arises for lower values of noise than the Hopf bifurcation, consistently, for all coupling strengths. For a fixed value of coupling allowing noise-induced oscillations, the oscillations will emerge from the clamped regime, as noise is increased, through the homoclinic bifurcation, and disappear through a Hopf bifurcation leading to the \emph{asynchrony} regime as noise is further increased.

\subsubsection{Noise-induced oscillations in a two-populations Wilson-Cowan equation} 

We conclude our analysis of universality with the study of a firing-rate model shown to exhibit noise-induced oscillations~\cite{touboul-hermann-faugeras:11}, and for which one can access rigorously the bifurcation diagram for the mean-field solutions. This model describes the activity of an excitatory (E) and an inhibitory (I) neuron populations of size $n_e$ and $n_i$:
\[\begin{cases}
	dx^i_t &= \Big(-x^i+\frac{J_{ee}}{n_e} \sum_{j=1}^{n_e} S(x^j)\\
	&\qquad \quad +\frac{J_{ei}}{n_i} \sum_{j=1}^{n_i} S(y^j)+I_e\Big)\,dt
	+\sigma dW^i_t\\                                                                                               
	dy^i_t &= \Big(-y^i+\frac{J_{ie}}{n_e} \sum_{j=1}^{n_e} S(x^j)\\
	&\qquad \quad +\frac{J_{ii}}{n_i} \sum_{j=1}^{n_i} S(y^j)+I_i\Big)\,dt + \sigma d\tilde{W}^i_t,
\end{cases}\]
where $I_e$ (respectively, $I_i$) is the deterministic level of current received by excitatory (respectively, inhibitory) cells and $W^i$ (respectively, $\tilde{W}^i$) are independent Brownian motions accounting for current fluctuations. Cells are coupled through the product of a nonlinear sigmoidal (smooth) transform of each neuron's activity multiplied by a coupling coefficient $J_{\alpha\beta}$ for $\alpha, \beta \in \{e,i\}$ representing the typical coupling strength of neurons of population $\beta$ onto neurons of population $\alpha$; these coefficients are assumed to be equal to $J_{\alpha\beta}=J g_{\alpha\beta}$ for a fixed connectivity matrix $G=(g_{\alpha\beta})_{\alpha\beta}$ and where $J$ acts as a scaling coefficient as in the previous cases. As shown in~\cite{touboul-hermann-faugeras:11}, this system converges to the mean-field equations:
\[\begin{cases}
	dx_t &= \Big(-x_t+J_{ee}\E[S(x_t)]+J_{ei} \E[S(y_t)]\Big)\,dt +\sigma dW_t\\
	dy_t &= \Big(-y_t+J_{ie} \E[S(x_t)]+J_{ii} \E[S(y_t)]\Big)\,dt+\sigma d\tilde{W}_t,
\end{cases}\]
which are implicit stochastic differential equations with dynamics coupled to the mean of the solution. These are generally complex mathematical equations to handle, but this particular case enjoys a massive simplification. Indeed, the coupling terms in this limit are deterministic expectations, and thus because of the Gaussian nature of the noise and linearity of the intrinsic dynamics, solutions to these stochastic equations are asymptotically Gaussian (or Gaussian for all times if the initial condition is), and its moments satisfy the ODE:
\[\begin{cases}
	\dot\mu_e &= -\mu_e+J_{ee} f(\mu_e,v)+J_{ei} f(\mu_i,v)\\
	\dot\mu_i &= -\mu_i+J_{ie} f(\mu_e,v)+J_{ii} f(\mu_i,v)\\
	\dot v &= -2 v+\sigma^2,
\end{cases}\]
with $f(\mu,v)=\int_{\R} S(x)e^{-(x-\mu)^2/2v}/\sqrt{2\pi v}\,dx$, a function with an explicit expression when  $S(x)$ is an error function (the repartition function of the Gaussian). This simplification thus allows addressing the existence of transitions in the collective dynamics using classical bifurcation theory for ODEs.

This system is not expressed as an excitable system, and does not have an explicit slow-fast structure. However, the geometry of its phase plane endows it with excitable properties. Indeed, a single pair of excitatory/inhibitory neurons in the absence of noise features a single stable fixed point, a saddle and an unstable spiral (see Fig.~\ref{fig:Universality}-C1). The unstable manifold of the saddle describes a large loop around the unstable fixed point (an heteroclinic orbit), and its stable manifold acts as an excitability threshold: perturbations of the fixed point across this manifold lead to a long excursion back to the stable fixed point along the unstable manifold heteroclinic orbit. We thus expect to find a similar transition to synchrony due to noise, provided that connections are synchronizing. Here, the excitatory/inhibitory nature of the system plays this role and maintains a cohesion in the system (although more indirectly than the electrical coupling). Indeed, when many excitatory neurons are activated, they will activate more excitatory cells, and when a majority is silent, other neurons tend to remain silent: therefore, excitatory neurons may act as the excitable voltage in FitzHugh-Nagumo or Morris-Lecar model, and inhibition may act as the adaptation variable. 

The mechanisms described in this paper thus account for the noise-induced oscillations observed in~\cite{touboul-hermann-faugeras:11}: for small noise, the system thus remains in the vicinity of the fixed point, but as noise increases, excursions through the stable manifold of the saddle will be more frequent, raising the number of excited neurons, and eventually leading to a collective spike. Transitions as a function of noise and a global connectivity parameter $J$ scaling all connectivity coefficients is provided in Fig.~\ref{fig:Universality}(C2), and confirms again the conjecture of the appearance of noise-induced oscillations through a homoclinic bifurcation at $\sigma$ small, and their disappearance through a Hopf bifurcation for $\sigma$ large.

\section{Discussion}

Noise is a prominent feature in natural or physical systems, arising from multiple sources including electrical fluctuations, thermal agitation of molecules or electrons and synaptic activity, to cite a few. In nonlinear systems, noise can have multiple effects, such as attractor switching in multistable systems, stochastic resonances in excitable systems~\cite{lindner:04} or inverse resonances near folds of limit cycles~\cite{gutkin,tuckwell2012analysis}. These phenomena have been largely studied for  finite-dimensional stochastic systems. The role  of noise in large-scale nonlinear interacting particle systems is significantly less well understood. A variety of articles have revealed the complex phenomenology noise may have in these systems, and particularly, a regularizing effect leading to the stabilization of stationary solutions or the emergence of periodic solutions~\cite{scheutzow1985noise,scheutzow1985some,zaks,lindner:04,lucon,quininao-touboul:18}. Despite increasing evidence and abstract mathematical proofs in the mean-field limit regime, the origin of these surprising regularizing effects of noise have remained elusive. 

In the first part of this paper, we thoroughly investigated this collective phenomenon of robust emergence of synchronized oscillations. Two essential microscopic mechanisms were highlighted underlying this behavior: asymmetric transitions from rest to an excited state and vice-versa, leading to an increase in the fraction of excited neurons, and a subsequent chain reaction recruiting all neurons into a collective excursion. Contrasting with other classical mean-field behaviors, these dynamics are therefore not driven by an influence of the ensemble average itself, but rather by the random and independent fluctuations of each neuron's activity, enabling the buildup of a macroscopic fraction of neurons in the  excited state. These phenomena are not specific to the FitzHugh-Nagumo model, but are universal to stochastic networks of excitable elements with confining interactions, as we confirmed in three other classical neural networks. 

The emergence of oscillations at unison arises in response to independent stochastic fluctuations of each element may seem surprising: one may have expected stronger correlations in the average activity in response to correlated inputs, or small noise. It is however an opposite effect that takes over: regular macroscopic trajectories emerge actually in response to having independent noise, making each neuron independent in the large network size limit, and thus the time to reach the critical fraction of excited neurons deterministic, much like an empirical average of independent variables converges to a deterministic value in the law of large numbers. Independent noise at the level of each neuron is a  relatively realistic assumption in the brain when accounting for the on-going bombardment of synaptic inputs  and channel fluctuations, and was supported by recent data~\cite{ecker-berens-etal:10,renart-de-la-rocha-etal:10}, and a general properties of large-scale stochastic networks (Boltzmann's molecular chaos hypothesis, or propagation of chaos).

This phenomenon shares a number of commonalities with brain activity disruptions occurring in Parkinson's disease. Indeed, in this disease, abnormal oscillations emerge and are associated with an increased excitability of cells~\cite{degos,degos2} and enhanced electrical  transmission~\cite{grace}. Motivated by this analogy, we next investigated the possible impact of high-frequency stimulation on the network, emulating DBS therapies. We observed that such a periodic forcing could prevent the emergence of synchrony. For this desynchronization to occur, stimulation have a frequency within a band of values, high relative to the frequency of intrinsic oscillations, but not too large. Low frequency stimulations forced the system to lock to the stimulus, very high frequency stimulations had no impact on the spontaneous oscillations. Between these two regimes, the system, poised near the chain reaction threshold, maximized the mutual information transmitted. 

Undoubtedly, our model does not constitute a realistic neural populations architecture and connectivity appropriate to reproduce closely the phenomena arising in Parkinson's disease. Despite this simplification, several parallels can be drawn between our model and clinical observations of parkinsonian patients~\cite{kuhn2008high,aum,lozano,ashkan} in particular, the dependence of the improvement in motor symptoms and the abolishment of beta oscillations (10-30 Hz) in stimulation frequency, restricted to high-frequency bands (130 Hz); the ratio between this effective stimulation frequency and the frequency of the intrinsic oscillatory rhythm is similar to what we observe in our model, both for the emergence of the desynchronized regime and maximal mutual information. Moreover, the amplitude of stimulation pulses can be optimized: our model indicates that a robust increase in mutual information can be found for relatively low amplitude stimulation, as soon as the desynchronization occurs. Therapeutically, low amplitude stimulation could seem beneficial since it avoids imposing high-amplitude currents that may be detrimental to network activity and may prevent the network from responding with a high signal-to-noise ratio to natural stimuli. In contrast to more realistic models of cortico-basal ganglia loops aiming at explaining the detailed mechanisms underlying the emergence of beta oscillations and the therapeutic effect of DBS~\cite{rubin2004high,Leblois3567}, we propose a more abstract line of thought linking macroscopic level to microscopic dynamics. In particular, our model proposes an alternative view to the ``information lesion'' hypothesis~\cite{grill}, suggesting that instead of producing highly regular output patterns, DBS endows the network with highly variable spontaneous activity, without any strong coherent activation of individual neurons, and thus high information processing capabilities. The statistical physics arguments we provided accounting for the loss of synchronization are, again, not specific to the particular model studied, and we argue that this phenomenon shall arise also in more realistic situations. In particular, the same type of protocols and background was used in a more realist model of the cortico-striatal loop in normal and parkinsonian conditions~\cite{charlotte}, and this work opens up a promising avenue for a better understanding of Parkinson's disease and its treatment. The study also highlights a number of open problems in mathematics: most phenomena described indeed challenge a well-developed theory of noise in nonlinear dynamics, and calls for extending these results beyond small noise and single elements. 

\textbf{Aknowledgements} The authors warmly thank Bertrand Degos and Marie Vandecasteele for multiple discussions on Parkinson's disease, as well as Valentin Figu\'e for his undergraduate research work in the early stages of the study.

\pagebreak
\section{Appendix}
\subsection{Transition rates and Kramer's theory}\label{sec:Kramer}
In the main text, we have derived numerically the transition rates from pioneer to rest and reciprocally through numerical simulations. These rates are analogous to transition rates of stochastic particles in multi-well potentials, and we discuss here the limitations of using the well-developed classical theory of attractor switching in multi-wells potentials to the problem  at hand.  

For a stochastic particle in a multi-well potential and in the limit of small noise, it is well-known that the particle switches attractor at exponentially distributed times with rates depending on the depth of the wells and the noise level~\cite{freidlin-wentzell:98,kramer}. Here, the problem at hand challenges the standard theory in many ways. One important difficulty stems from the fact that noise-induced oscillations arise for non-vanishing noise, a regime where little remains known about transitions between multiple attractors and where large-deviations estimates are ineffective. Moreover, each particle has an excitable dynamics instead of a multi-stable Hamiltonian dynamics: while rest is indeed an equilibrium, the pioneer state is a transient passage that do not correspond to a stable fixed point. 

One way around the latter difficulty is to reducing the system to a one-dimensional equation with $w$ a constant during the time of a given transition, as done multiple times in section~\ref{sec:NoiseInducedSync}. This allows deriving an effective potential to approximate system during the transition phase. Unfortunately, the potential obtained would now depend on the level of noise and of the distribution of the voltages of neurons. Indeed, equation~\eqref{eq:Particle}, corresponds to the dynamics of a noisy particle in a one-dimensional potential $U$ given by:
\begin{align*}
	&U(v) =-\int_0^v f(y)\,dy + J\frac{v^2}{2} + A v \\
	\quad &=\frac{v^4}{4} - \frac{(1+a)v^3}{3} + \frac{(J+a) v^2}{2}+ A v.
\end{align*}
with $A=(w_0 - J ((1-\alpha) v_r+\alpha v_p)$. This potential requires assuming that the system, during one transition, conserves a fixed value of $w_0$, a fixed fraction $\alpha$ of neurons in the vicinity of a pioneer state $v_p$ and the rest of neurons in the vicinity of the resting potential $v_r$. Both assumptions are reasonable for the system at hand, but two difficulties arise to obtain analytical results using this formula. First, the potential depends on noise and coupling through $w_0$. Moreover, $v_r$ and $v_p$ are also not fully defined and depend on $w_0$ (thus on noise and $J$) and $\alpha$. Indeed, $v_r$ and $v_p$ shall correspond to the resting and pioneer state, two putative minima of the potential. Computing these quantities thus amounts to solving an implicit equation:
\[f(v) - Jv + w_0 - J ((1-\alpha) v_r+\alpha v_p)=0\]
where $v_r$ is the smallest and $v_p$ the smallest solution of the cubic equation, when that equation has three solutions. These in particular only exist for a limited range of values of $(\alpha,\sigma,J)$ for which $U$ is a double-well potential. The dependence in $w_0$ of the potential highlights the interplay between noise and nonlinearity in the system, while the dependence in $\alpha$ shows the collective nature of the problem, both being fundamental differences with the classical attractor switching framework. 

Despite these differences, we computed, when possible, the effective potential of the system. To this purpose, we derived the values of $v_r,v_p$ using a fixed point method. In detail, fixing $v_r^0$ and $v_p^0$ as the smallest and largest solutions of $f(v)=w_0$, we iteratively solved (when possible) the equation:
\[f(v^{n+1}) - Jv^{n+1} + w_0 - J ((1-\alpha) v_r^{n}+\alpha v_p^n)=0\]
for $n\geq 0$ and checked for convergence of the sequence $(v_r^n,v_p^n)$. Convergence of that sequence ensured the fact that the effective potential has indeed a double-well profile. Denoting $v_u$ the voltage associated with the saddle (unstable fixed point), Kramer-Freidlin-Wentzell theory of attractor switching provides an asymptotic expansion of the escape time depending on the shape of the potential $U$ when $\sigma$ is small:
\begin{equation}\label{eq:Kramer}
	\frac{2 \pi}{\sqrt{-U''(v_u)U''(v^*)}} \exp\left(\frac{U(v_u)-U(v^*)}\sigma\right)
\end{equation}
where $v^*$ is the voltage of one of the stable nodes of the system ($v_p$ or $v_r$). 

\begin{figure}[h]
	\includegraphics[width=\columnwidth]{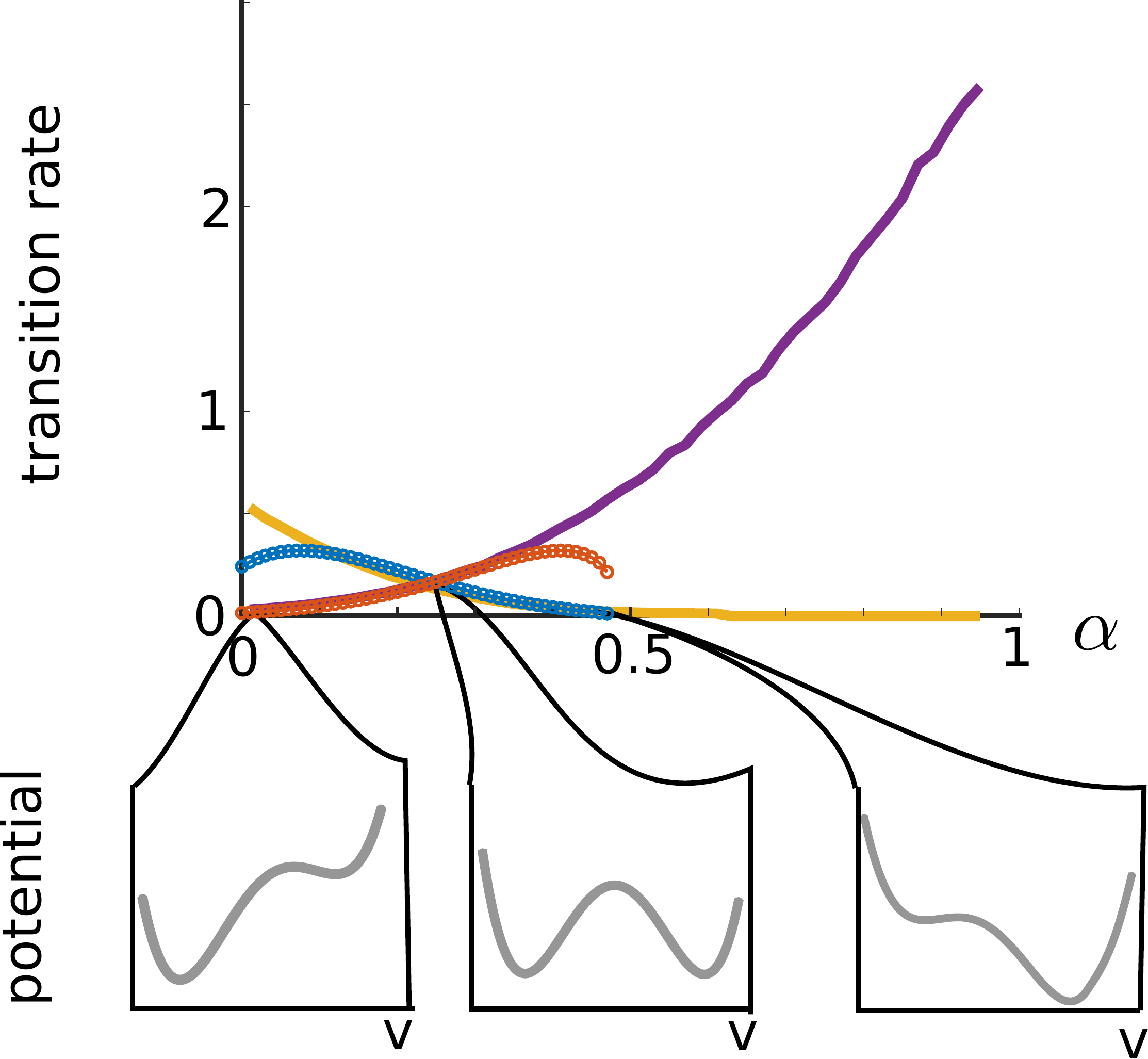}
	\caption{Comparison of numerical rates of transitions from pioneer to rest (yellow) or rest to pioneer (purple) with associated Kramer's escape rate formula~\eqref{eq:Kramer} (respectively, blue and red) for $J=1.5$ and $\sigma=1.5$. A very good agreement arises when the equilibria are sufficiently stable (potential wells of $U$ deep enough). Escape rates are largely under-estimated when the potential becomes too flat (conjectured breakdown arising when the depth becomes of the same order of magnitude as the potential well depth), in which case typical stochastic fluctuation dominate the escape rate compared to the small-noise correction provided by the theory.}
	\label{fig:Kramer}
	\end{figure}

Figure~\ref{fig:Kramer} compares the rates provided by the Kramer-Freidlin-Wentzell theory with those obtained numerically in section~\ref{sec:Transitions}. We found an excellent agreement of the rates where it can be computed (presence of a double-well potential) and when the wells are deep enough. In detail, rest and pioneer states for the potential can be defined for a finite interval of values $\alpha$; for $\alpha$ too small, the pioneer state is barely attractive to a particle starting from the resting regime, and for $\alpha$ too large, the resting state becomes progressively less stable and even disappears. In flat regions of the potential, the analytical rate deviates from the numerical rate computed and over-estimates the time it will take for a particle to cross the separatrix. Indeed, a noisy particle in a potential whose depth is of the same order of magnitude as typical amplitude of stochastic excursions will have a larger transition rate than predicted by the theoretical asymptotic escape rate in small noise. The numerical method used in section~\ref{sec:Transitions} avoids these difficulties and computes the effective times of transitions regardless of the assumptions needed for applying Kramer-Freidlin-Wentzell theory.

	\captionsetup[figure]{name={Supplementary Figure S},labelsep=period}

	\captionsetup[figure]{name={Movie M},labelsep=period}
	\setcounter{figure}{0}
	
	\begin{figure}
		\caption{Trajectories of a subset of $100$ neurons in a network of $n=4\,000$ neurons in the phase plane for each of the 5 parameter set shown in Fig.~\ref{fig:phenomenon} A. Blue curve: $v$-nullcline, Red: $w$-nullcline. Yellow circles: resting neurons, Cyan circles: pioneers/ excited neurons.} 
		\label{mov:m1}
		\end{figure}
	
\end{document}